\documentclass[a4paper]{article}

\usepackage{jcappub}

\usepackage{aasmacros}

\title{\boldmath Nonlinear spherical perturbations in Quintessence
  Models of Dark Energy}

\author[a]{Manvendra Pratap Rajvanshi,}
\author[a,1]{J. S. Bagla,\note{Corresponding author}}

\affiliation[a]{IISER Mohali, Sector 81, Sahibzada Ajit Singh Nagar,
  Punjab 140306, India} 

\emailAdd{manvendra@iisermohali.ac.in}
\emailAdd{jasjeet@iisermohali.ac.in}

\abstract{Observations have confirmed the accelerated expansion of the
universe.
The accelerated expansion can be modelled by invoking a cosmological
constant or a dynamical model of dark energy.
A key difference between these models is that the equation of state
parameter $w$ for dark energy differs from $-1$ in dynamical dark
energy (DDE) models.
Further, the equation of state parameter is not constant for a general
DDE model.  
Such differences can be probed using the variation of scale factor
with time by measuring distances.
Another significant difference between the cosmological constant and
DDE models is that the latter must cluster.
Linear perturbation analysis indicates that perturbations in
quintessence models of dark energy do not grow to have a significant
amplitude at small length scales. 
In this paper we study the response of quintessence dark energy to
non-linear perturbations in dark matter.
We use a fully relativistic model for spherically symmetric
perturbations.
In this study we focus on thawing models.
We find that in response to non-linear perturbations in dark matter,
dark energy perturbations grow at a faster rate than expected in
linear perturbation theory. 
We find that dark energy perturbation remains localised and does
not diffuse out to larger scales.
The dominant drivers of the evolution of dark energy perturbations are
the local Hubble flow and a supression of gradients of the scalar
field. 
We also find that the equation of state parameter $w$ changes in
response to perturbations in dark matter such that it also becomes a
function of position.
The variation of $w$ in space is correlated with density contrast for
matter. 
Variation of $w$ and perturbations in dark energy are more pronounced
in response to large scale perturbations in matter while the
dependence on the amplitude of matter perturbations is much weaker. 
}

\begin{document}

\maketitle

\section{Introduction}
\label{sec:intro}

One of the central themes of modern cosmology is the quest for
understanding the source of the observed acceleration of the expansion
rate of the Universe.  
Early evidence for an accelerating universe came from clustering of
galaxies\cite{1990Natur.348..705E}.
There were several independent observations indicating that the
density parameter for matter is well below
unity\cite{1993Natur.366..429W,1995Natur.377..600O,1996ComAp..18..275B}.   
The observations of high redshift supernovae of type Ia ruled out a
universe without dark
energy\cite{1998ApJ...507...46S,1998AJ....116.1009R,1999ApJ...517..565P}. 
Observations of the cosmic microwave background radiation temperature
anisotropies require the total density parameter to be very close to
unity\cite{2000ApJ...536L..63M,2003ApJS..148..175S}, thus the universe
is a mix of normal matter, radiation, dark matter and dark
energy\cite{2016AA...594A..13P}.
Latest constraints on dark energy models and some discussion around
the origin of these constraints may be found in
\cite{2010MNRAS.405.2639J,2014MNRAS.441.3643P,2015MNRAS.446.1321H,2017arXiv171000846J,c577c8168f6e4d9c8cc679edcd67e494,2017JCAP...07..040D,2017JCAP...06..012T,2018arXiv180108553V}. 

A number of theoretical models have been proposed to explain the
accelerated expansion of the universe.
The simplest model that is consistent with observations is the so
called cosmological constant $\Lambda$, this suffers from the problem
of fine tuning\cite{1989RvMP...61....1W,2008GReGr..40..529P}.  
A number of dynamical dark energy models like Quintessence, k-essence,
chaplygin gas, etc. have been proposed, for an overview see the book
{\sl Dark Energy: Theory and Observations}\cite{2010deto.book.....A}. 
There are many models that rely on modified theories of gravity.
We refer the readers to recent reviews for more details on models of
dark energy and observational
constraints\cite{2008GReGr..40..529P, 2012ApSS.342..155B,
  2013CQGra..30u4003T,2013PhR...530...87W,2018RPPh...81a6901H}.     

Given the large number of possible explanations, the task is to
constrain these using observations and also rule out some
possibilities.
However, all the models can be {\sl tuned} to produce almost any
specified evolution of the scale factor\cite{2002PhRvD..66b1301P}.
Thus it is not possible to distinguish between different classes of
models using only the evolution of the scale factor.
It has been pointed
out that the evolution of perturbations in dark 
energy may be used to distinguish between different classes of
models\cite{2009PhRvD..79l7301J}.
Perturbative analysis shows that the growth of perturbations in dark
energy at small length scales is strongly suppressed, whereas dark
energy perturbations can develop a comparable amplitude to
perturbations in matter at very large length
scales\cite{2008PhRvD..78l3504U}. 

In this paper we study the evolution of dark energy perturbations at
small scales.
Our aim is to study the response of dark energy to non-linear
perturbations in matter, and to check whether perturbations in dark
energy have any discernible effect on dark matter perturbations. 
In order to simplify analysis, we restrict ourselves to spherically
symmetric perturbations. 

The spherical collapse model was introduced by Gunn \&
Gott\cite{1972ApJ...176....1G} where they used it to make a
theoretical connection with observational properties of Coma cluster.
The spherical collapse model has been used very successfully to
understand many aspects of structure formation.
A mapping between the linear and non-linear collapse is used in the
theory of mass function of collapsed halos and its generalisations.
Thus a detailed study of spherical collapse for any dark energy model
has many potential applications. 

Dynamics in a model with a cosmological constant has been studied by
many authors, including studies of spherical
collapse\cite{1984ApJ...284..439P,1996MNRAS.282..263E,1991MNRAS.251..128L,1993MNRAS.262..717B}. 
Barrow \& Saich \cite{1993MNRAS.262..717B} generalised the spherical
collapse model to include the cosmological constant.
In this case the equations can be reduced to quadrature and expressed
in terms of elliptic integrals.
They found that the presence of dark energy leads to a competition
between attractive gravity and repulsive dark energy, and small
density perturbations do not collapse.
Over densities need to be higher than a threshold if these are to form
a collapsed object.
Perturbations take longer time to collapse, collapsed perturbations
are larger for the given mass and hence have a lower density as
compared to perturbations in the Einstein-deSitter
model.  

In this article we present results from study of spherical collapse in
a cosmology with dark energy modeled by a minimally coupled canonical
scalar field called 'Quintessence'\cite{1998PhRvL..80.1582C} (also see
\cite{2013CQGra..30u4003T}).  
While there have been earlier attempts at modelling spherical collapse
of matter with Quintessence or other models of dark energy, e.g., see
\cite{2007JCAP...11..012A,2009PhRvD..79b3516A,2014PhRvD..89h3002R,2010JCAP...03..027C,2010JCAP...10..014B,2011JCAP...11..014A,2012JCAP...01..025M,2015AASP....5...51T,1475-7516-2016-09-031,PhysRevD.93.043533,PhysRevD.95.064029,0004-637X-841-1-63,2017PhRvD..96h3506A,2018PDU....19...12C},
almost all of these 
employ either some perturbative approximation scheme, or make a
strong assumption about quintessence field like non-clustering, zero
speed of sound, etc. 
In this study, we do not make any assumption/approximation for field
or metric apart from  spherical symmetry.
We consider fully non-linear, relativistic dynamics of space-time,
matter and scalar field. 
In \S{\ref{sec:scinq}}, we introduce the formalism and equations.
Initial conditions are presented in \S{\ref{subsub:ini}}, while
virialisation conditions and approach after virialisation is described
in \S{\ref{subsec:vir}}.  The discussion of results
(\S{\ref{sec:Results}}) is organized as follows:
\S{\ref{subsec:dmp}} deals with dark matter perturbations, whereas
dark energy/scalar field perturbation are discussed in
\S{\ref{subsec:DEP}}. 
Finally we highlight important results along with a discussion of their
implications in \S{\ref{sec:summary}}. 

\section{Spherical Collapse}
\label{sec:scinq}

In this section we outline our model in terms of equations.
The spherical collapse of non-relativistic matter, {\it aka}\/ dust
can be studied using a metric with spherical symmetry.
In case of a universe with any combination of matter, curvature and
dust, it can be shown that the Newtonian limit is exact.
We are dealing with non-relativistic matter and a scalar field, and in
this case there is no appropriate Newtonian limit.
Hence we have to work with a fully relativistic model in order to
provide a self consistent treatment for the combination of dust and
the scalar field. 

For modelling spatially isotropic perturbations, we start by
considering a general spatially isotropic metric in comoving
frame\cite{1934PNAS...20..169T,1947MNRAS.107..410B}: 
\begin{equation}
ds^2\,=\,-e^{(2B)}dr^2 - R^2(d\theta^2 + sin^2\theta d\phi^2) +
dt^2      \label{eq:Q1} 
\end{equation}
where $B(t,r)$ and $R(t,r)$ are arbitrary functions of $r$ and $t$.
Some of the characteristics of the metric in presence of pressure are
discussed by Lynden-Bell, D. and Bi{\v c}{\'a}k,
J.\cite{2016CQGra..33g5001L}. 
We have to solve for these two functions by solving the Einstein's
equations along with field equations and equations governing the
evolution of matter density.
The full set of equations is as follows:
\begin{eqnarray}
\ddot{B} &=& -c^2e^{-2B}\frac{R'^2}{R^2} + \frac{c^2}{R^2} +
\frac{\dot{R}^2}{R^2} - \dot{B}^2 - 4\pi G\rho - \frac{8\pi G}{c}
\left[ \frac{\dot{\psi}^2}{2c^2}  - e^{-2B}\frac{\psi'^2}{2} \right]  
\label{eq:Q26} \\
\frac{\ddot{R}}{R} &=& -\frac{4\pi G}{c}\left[
  \frac{\dot{\psi}^2}{2c^2} + \frac{e^{-2B}\psi'^2}{2} - V  \right]
-\frac{1}{2}\frac{\dot{R}^2}{R^2} + \frac{c^2}{2} \left[ 
  e^{-2B}\frac{R'^2}{R^2} - \frac{1}{R^2} \right]   \label{eq:Q27} \\
\ddot{\psi} &=& c^2\left[-\frac{\partial V}{\partial \psi} + e^{-2B}
  \left\lbrace \psi^{''}  - \left( B'-\frac{2R'}{R} \right)\psi'
  \right\rbrace \right] - \left(\dot{B}+\frac{2\dot{R}}{R}\right)
\dot{\psi} \label{eq:Q14} \\
\dot{\rho_m} &=& -\left( \dot{B} + \frac{2\dot{R}}{R} \right)\rho_m
\label{eq:Q25}
\end{eqnarray}
Here a dash represents a partial derivative with respect to $r$ and a
dot represents a partial derivative with respect to $t$.
In this problem, these are the two independent variables. 
In the present study we work with two potentials($V\propto \psi^2$
and $V\propto \exp(-\psi)$). 

The structure of the equations is amenable to defining the initial
conditions for the variables $B$, $R$, $\dot{B}$, $\dot{R}$, $\rho_m$,
$\phi$ and $\dot\phi$ at all $r$ and then evolving the system. 
We use a RK-4 based numerical scheme to solve these equations.
See Appendix \ref{Numerix} for details.

\subsubsection{Initial conditions}
\label{subsub:ini}

We first solve the equations in absence of any perturbations.
In this case there is no dependence on $r$ and the system is identical
to a FLRW
universe\cite{1922ZPhy...10..377F,1931MNRAS..91..483L,1935ApJ....82..284R,1935QJMat...6...81W}.  
In the FLRW limit $B(t,r)\rightarrow log(a(t))$ and $R(r,t)\rightarrow
a(t)r$. 
We require the universe to have $30\%$ dust or non-relativistic
matter, and $70\%$ dark energy at the present epoch.
The latter is contributed by the scalar field.
Further, we require $w$, the effective equation of state to be close
to $-1$. 
We also use an additional assumption that $\dot\psi=0$ at the initial
time. 
These requirements allow us to fix all unknown parameters related to
the scalar field.
We set the initial conditions at $z \sim 1000$.

\begin{figure}[tbp]
\centering 
\includegraphics[width=.45\textwidth]{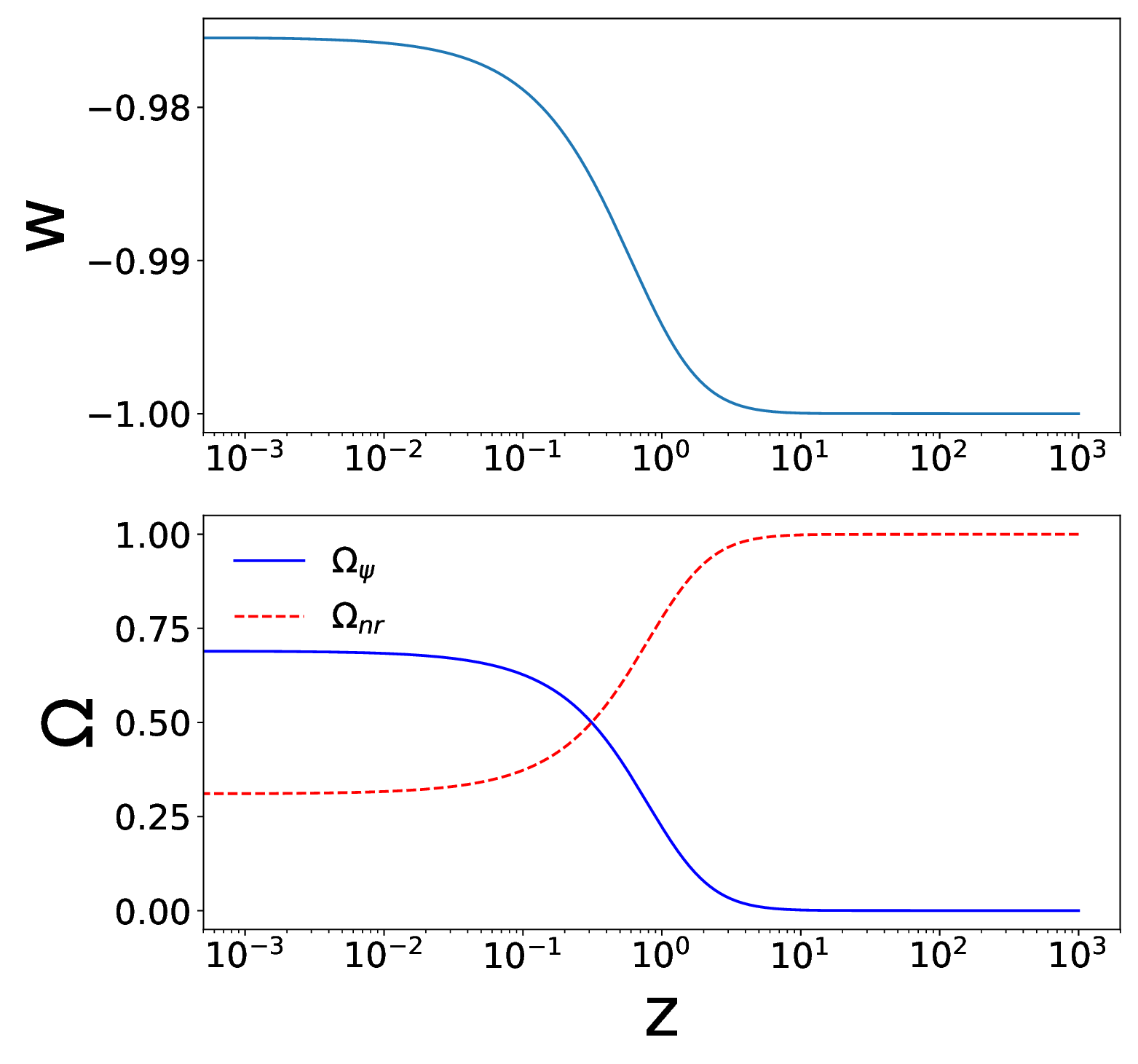}
\includegraphics[width=.45\textwidth]{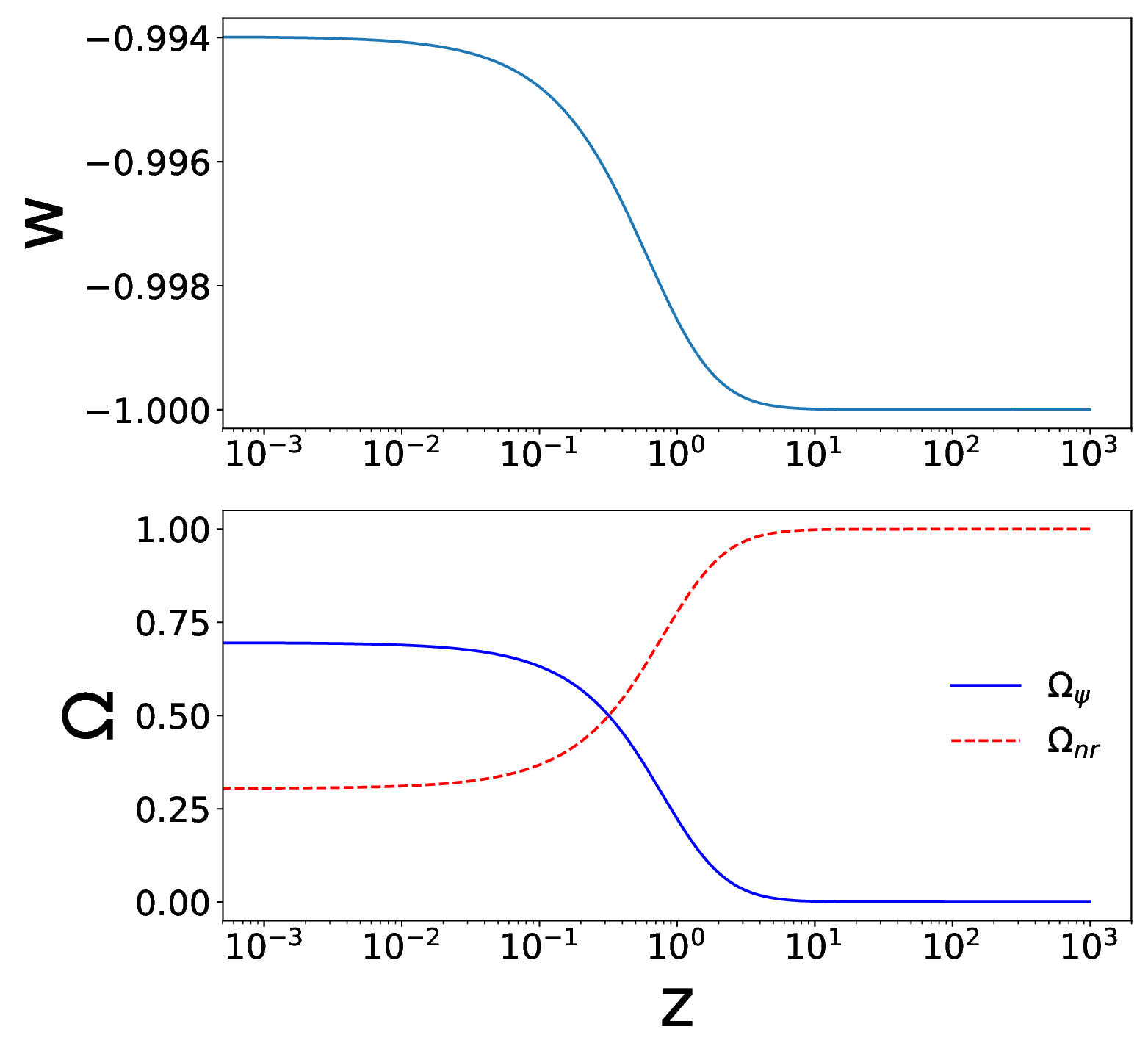}
\caption{\label{fig:1}  Evolution of the model without any
  perturbations.  This is shown for the two potentials: $V\propto
  \psi^2 $ (left column) and $V\propto \exp(-\psi) $ (right column).
  The top row shows the evolution of the equation of state parameter
  $w$ as a function of redshift $z$.  The lower row shows the
  evolution of the density 
  parameter for non-relativistic matter and the dark energy
  components.  We use initial conditions for the scalar field used
  here in all the simulations presented here unless mentioned
  otherwise.  Initial conditions for dark matter at large $r$ also
  fall back to this set.}    
\end{figure}

We first solve for evolution of the scale factor.  
We start the field with zero kinetic energy, i.e., $w=-1$.
We see that for both potentials being considered here, we can get
solutions where the equation of state parameter stays fairly close to
$w=-1$ up to the present time.
We have shown the evolution of $w$ and the density parameter for
matter ($\Omega_{nr}$) and field ($\Omega_{\psi}$) in
figure~\ref{fig:1}.  
This exercise allows us to set the parameters of the scalar field at
the initial time for the case where we study non-linear evolution of
perturbations. 

We set the initial conditions for the case with perturbations at $z
\sim 10^3$.
We assume that the scalar field representing dark energy is uniform at
this time.
This has been shown to lead to the expected adiabatic mode for
quintessence models \cite{2008PhRvD..78l3504U}.
This has also been noted by other authors who have studied attractors
for dark energy perturbations \cite{2010JCAP...10..014B}.
We study results of our calculations at late times, $z \leq 10$ and
hence there is adequate time for the solution to approach the
attractor. 
As mentioned above, the scalar field initial conditions are set by
assuming that there are no perturbations and we start with $w=-1$: 
\begin{eqnarray}
  \psi_i' &=& 0 =  \psi_i'' \nonumber \\
  \dot{\psi}_i &=& 0 \nonumber \\
  \psi_i &=& 1
\end{eqnarray}
Functional form of potentials have some parameters like amplitude of
potential $V_0$ and $\lambda$ in $V=V_0\exp(-\lambda \psi)$.
These parameters and initial values are chosen to get the
desired background evolution.

The matter distribution has a small initial perturbation.
The profile of matter perturbation is compensated, i.e, the central
perturbation is offset by a perturbation of opposite sign so that at
large $r$, the net perturbation integrated over volume goes to zero.
We use the following functional form:
\begin{equation}
\delta_i(r) = \left\{ \begin{array}{lll} \alpha_0
  \left[1-\left(\frac{r}{\sigma_0}\right)^2 
  \right]^2 -\alpha_1  \left[1-\left(\frac{r}{\sigma_1}\right)^2
  \right]^2 & {\,\,\,\,} & \left(r\leq\sigma_0\right)  \\
 -\alpha_1  \left[1-\left(\frac{r}{\sigma_1}\right)^2 \right]^2 & {\,\,\,\,} &
 \left(\sigma_0 < r\leq\sigma_1\right) \\
 0   & {\,\,\,\,} & \left(r>\sigma_1\right) \\
\end{array}
\right.
\end{equation}
Here we require $\sigma_0 < \sigma_1$.
The requirement of net perturbation after averaging over volume to
$r=\sigma_1$ can be stated as:
\begin{equation}
\int\limits_0^{\sigma_1} \delta_i(r)r^2 dr = 0  
\end{equation}
Thus there is no net perturbation at scales larger than $\sigma_1$ and
these regions should evolve as a smooth universe.  
This leads to the following relation between $\alpha_0$ and $\alpha_1$: 
\begin{equation}
\alpha_1 = \alpha_0\left( \frac{\sigma_0}{\sigma_1} \right)^3
\end{equation}

We set initial velocity of each shell by assuming that these are
comoving with the uniform Hubble expansion.
This facilitates comparison as this assumption has been used in
earlier studies \cite{1993MNRAS.262..717B} \cite{1972ApJ...176....1G}
as well. 
In comparison with linear theory it is important to recall that only
$3/5$ of the initial density perturbation in such a case is in the
growing mode\footnote{While the number $3/5$ is derived for the
  Einstein-deSitter model, it is a useful approximation as much of the
evolution takes place in the matter dominated phase.}.
Using this with initial condition $R_i=a_i r$, we can obtain initial
conditions for metric coefficients and their time derivatives.
For numerical convenience, we redefine the time variable as
$t\rightarrow tH_i$ where $H_i$ is initial value of the Hubble
parameter.  
\begin{eqnarray}
B_i &=& \ln(a_i)-
        \frac{1}{2}\ln\left[1-
        \frac{3}{r}\frac{\Omega_{im}a_i^2}{c^2}\int  
        dr r^2\delta(r)  \right]\\ 
\dot{B}_i &=& 1\\
R_i &=& a_i r\\
{\dot{R}}_i &=& R_i \\
R' &=& a_i \\
R'' &=& 0 
\end{eqnarray} 
The subscript $i$ refers to the initial value of the variable, $a$ is 
the scale factor and $H$ is the Hubble parameter.
$\Omega_{im}$ is the initial value of density parameter for matter. 

The argument of the logarithm in the second term in the expression for
$B_i$ must be positive and hence there is a restriction of the
amplitude and scale of perturbations we can simulate.
In particular this affects simulations of large scale over densities:
comoving initial conditions for arbitrarily large perturbations are
not allowed. 

The generic solution for a shell with over density is that it expands
with the universe at early times.
The expansion rate slows down as the gravitational pull of excess mass
leads to a more rapid deceleration.
The shell reaches a maximum radius, also known as turn around.
This is followed by a collapsing phase where the shell falls towards
the centre. 

\subsection{Virialisation and beyond}
\label{subsec:vir}

Mathematical solutions to general spherical collapse lead to formation
of a singularity as each shell with a sufficiently high over density
collapses to the origin.  
In a realistic scenario, velocity dispersion as well as non-radial
motions that are negligible in the early phase dominate at late
times. 
It is also expected that violent relaxation will drive the system
towards virial equilibrium.
These are expected to play an important role for dark matter as it
cannot radiate or loose energy via any other channel.
We proceed with a simplistic approach assuming that in-falling
perturbation stabilizes at radius where kinetic energy and potential
energy satisfy virial theorem \eqref{eq:Q29}.
In case of Einstein-deSitter universe, this leads to a simple
expression for the virial radius: the radius of the virialised halo is
exactly half of the maximum or the turn around radius for the shell
\citep{1972ApJ...176....1G}. 
Barrow and Saich\cite{1993MNRAS.262..717B} generalised this
to the case when a non-zero cosmological constant is present besides
non-relativistic matter. 
\begin{equation}
\begin{split}        \label{eq:Q291}
R_V = &\left(\frac{2}{3}\right)^{1/3} \left( \frac{\Omega_\Lambda
  R_T^3+\Omega_M (\frac{a_0}{a_{in}})^3 (1+\delta_{in})
  R_{in}^3}{\Omega_\Lambda R_T}  \right)^{1/2}  \\ & \sin\left[
  \frac{1}{3} \arcsin \left\{ \frac{\Omega_M 
    a_0^3 (1+\delta_{in}) R_{in}^3}{a_{in}^3 R_T^3}
   \left( \frac{1.5}{1  + \frac{\Omega_M}{\Omega_\Lambda} (\frac{a_0
       R_{in}}{a_{in} R_T})^3 (1+ \delta_{in}) }
 \right)^{3/2}\right\} \right]   
\end{split}
\end{equation}   
Here, $R_{in}$ is the initial radius, $R_T$ is the maximum or the turn
around radius, $\delta_{in}$ is the initial density contrast inside
the shell, $a_{in}$ is the initial value of the scale factor and $a_0$
is its present value, $\Omega_\Lambda$ is the density parameter
corresponding to the cosmological constant at present and $\Omega_M$
is the density parameter for non-relativistic matter.

Calculations are much harder in the case of dynamical dark energy.
Maor and Lahav \cite{2005JCAP...07..003M} summarize two limiting cases
for a fluid model of dark energy. 
They point out that there are significant differences that arise
depending upon whether or not dark energy participates in the
virialisation process. 
The two limiting cases they consider are: only dark matter virialises
and dark energy does not cluster, and, both dark matter and dark
energy virialise. 
Maor and Lahav \cite{2005JCAP...07..003M} show that if only dark
matter virialises, then the ratio of virial radius to turn around
radius is on lower side of Einstein-DeSitter value of $0.5$, while if 
the two component system of dark energy plus dark matter virialises
together, then this ratio is larger than half.
It is relevant to note here that in the case of a cosmological
constant, the expected ratio of virial radius to turn around radius is
less than half. 

As we shall see below, we find that in the case of scalar field, the
ratio of virial radius to the turn around radius is less than half.

\subsubsection{Evolution of dark energy beyond virialisation}
\label{subsec:aftervir}

We use the Virialisation condition:
\begin{equation}
<T> + \frac{1}{2}\left\langle R \, F_{R} \right\rangle = 0 \label{eq:Q29}
\end{equation}
here $T$ is the kinetic energy, $R$ is the radius of the shell and
$F_R$ is the radial force on the shell.
Angular brackets denote averaging over time.
\begin{eqnarray}
T=\frac{1}{2}\dot{R}^2 \quad F_{R} = \ddot{R}  \label{eq:Q30}
\end{eqnarray}
In case of cosmological constant one can use this relation to get an
analytical form \eqref{eq:Q291} for $R_{V}$ is terms of $R_{T}$
\citep{1993MNRAS.262..717B}. 
In case of quintessence being considered here, we track the value of
the right hand side of Eqn.\ref{eq:Q29} after turn around and declare
the shell to have virialised when this value becomes zero for the
first time. 

It is pertinent to note that this implicitly takes the time of
virialisation to be the time when the shell reaches virial radius
during the collapsing phase.
This is different from the usual interpretation where it is assumed
that virialisation happens at the time when the shell collapses to the
origin.
An implication of this is that the density contrast at the time of
virialisation computed here is lower than that obtained with the usual
method as the background density is higher.
For reference, note that in case of an Einstein-deSitter background,
the density contrast at virialisation with this approach is $145$, as
compared to $168$ that we obtain using the usual method.

After turn around, we check for condition \eqref{eq:Q29} and at that
particular $R(r)$ we freeze the metric terms $B(t,r)$ and $R(t,r)$,
and we do so because $R(t,r)$ has physical meaning of physical radius
which stabilizes at virialisation.
In case of $B(t,r)$ we take a cue from $\Lambda CDM$ where $B(t,r)$ is
dependent on spatial derivatives of $R(t,r)$.
Further, consistency requires that we set time derivatives of the two
variables to zero.

As we freeze the metric coefficients, the set of equations we have can
no longer be evolved self consistently.
Therefore the solutions at later times, after virialisation of the
innermost shells, are approximate solutions. 
As we shall see, dark matter dominates over dark energy in the
virialised region and hence an approximate solution can be attempted
without expecting a significant back reaction and an implied variation
of metric coefficients. 
The scalar field equations need to be solved over the entire range of
scales and it is not obvious whether any choices we make for the
solution in the interior of the virialised region will have an impact
on the evolution of the field at large scales.

We try three approaches to approximate solution for the scalar field
in the virialised region.
\begin{enumerate}
\item
  The scalar field can be evolved as a test field in the space-time
  determined by the frozen metric coefficients in the virialised
  region.
\item
  The scalar field can also be frozen in the virialised region, i.e.,
  we put $\dot{\psi} = 0 = \ddot{\psi}$ in this region.
\item
  We put $\ddot\psi=0$ and freeze the value of $\dot\psi(r)$ inside
  the virial region.
\end{enumerate}

We compare the three approaches and show that these lead to similar
evolution at scales around the virial radius.
Further, we show that the solutions are indistinguishable at scales
larger than the turn around radius.

In the first approach given above, we solve for the scalar field
inside the virial radius according to the following equation:
\begin{equation}
\label{eq:Q31}
\ddot{\psi} = c^2\left[-\frac{\partial V}{\partial \psi} +
  e^{-2B_{vir}} \left\lbrace \psi^{''}  - \left(
      B'_{vir}-\frac{2R'_{vir}}{R_{vir}} \right)\psi' \right\rbrace
\right]  
\end{equation}  
Here, $R_{vir}$ and $B_{vir}$ are the frozen values of metric
coefficients inside the virial radius.
We solve the full set of equations outside the virial radius.

\begin{figure}[tbp]
\centering 
\includegraphics[width=.45\textwidth]{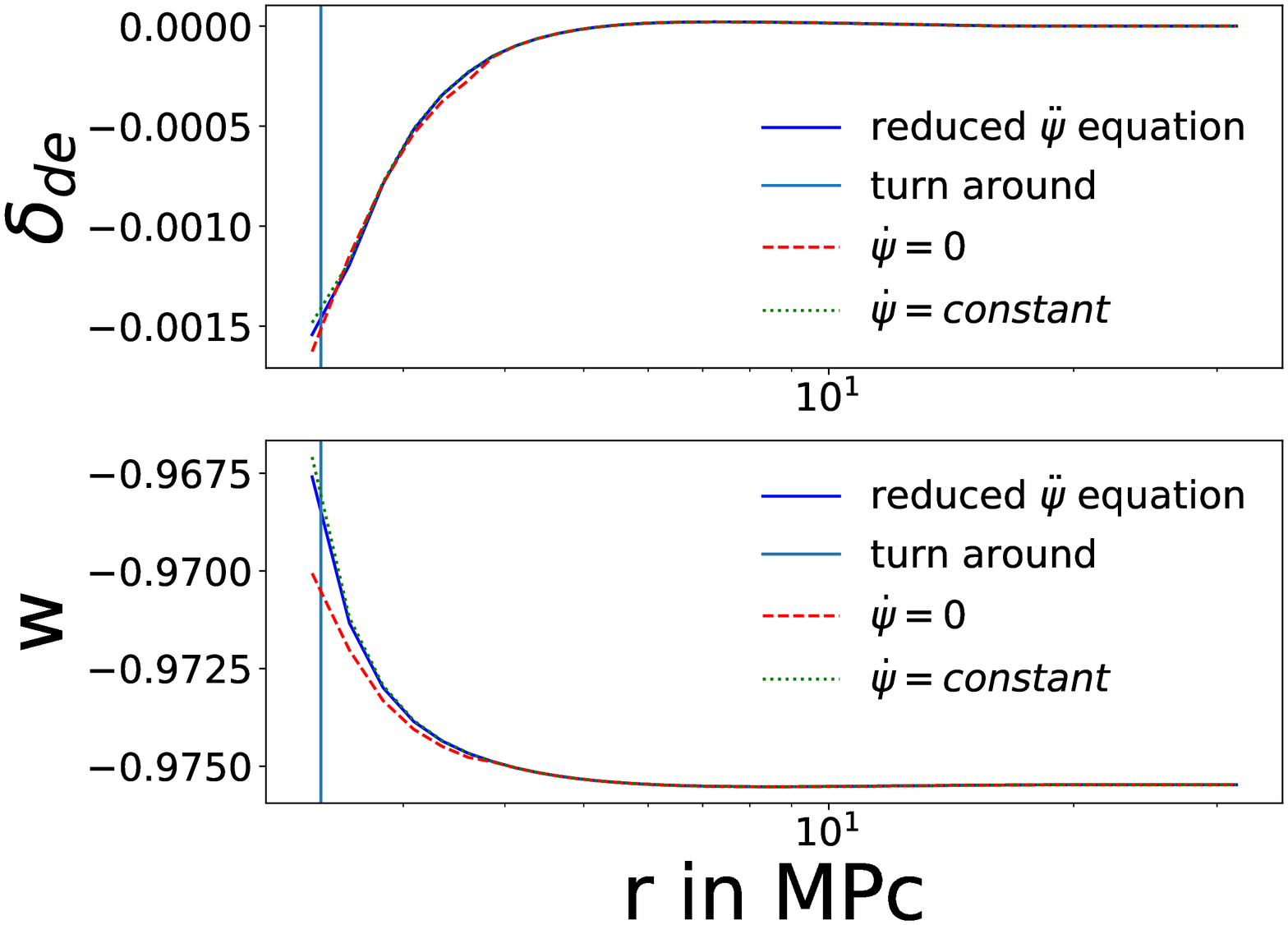}
\includegraphics[width=.45\textwidth]{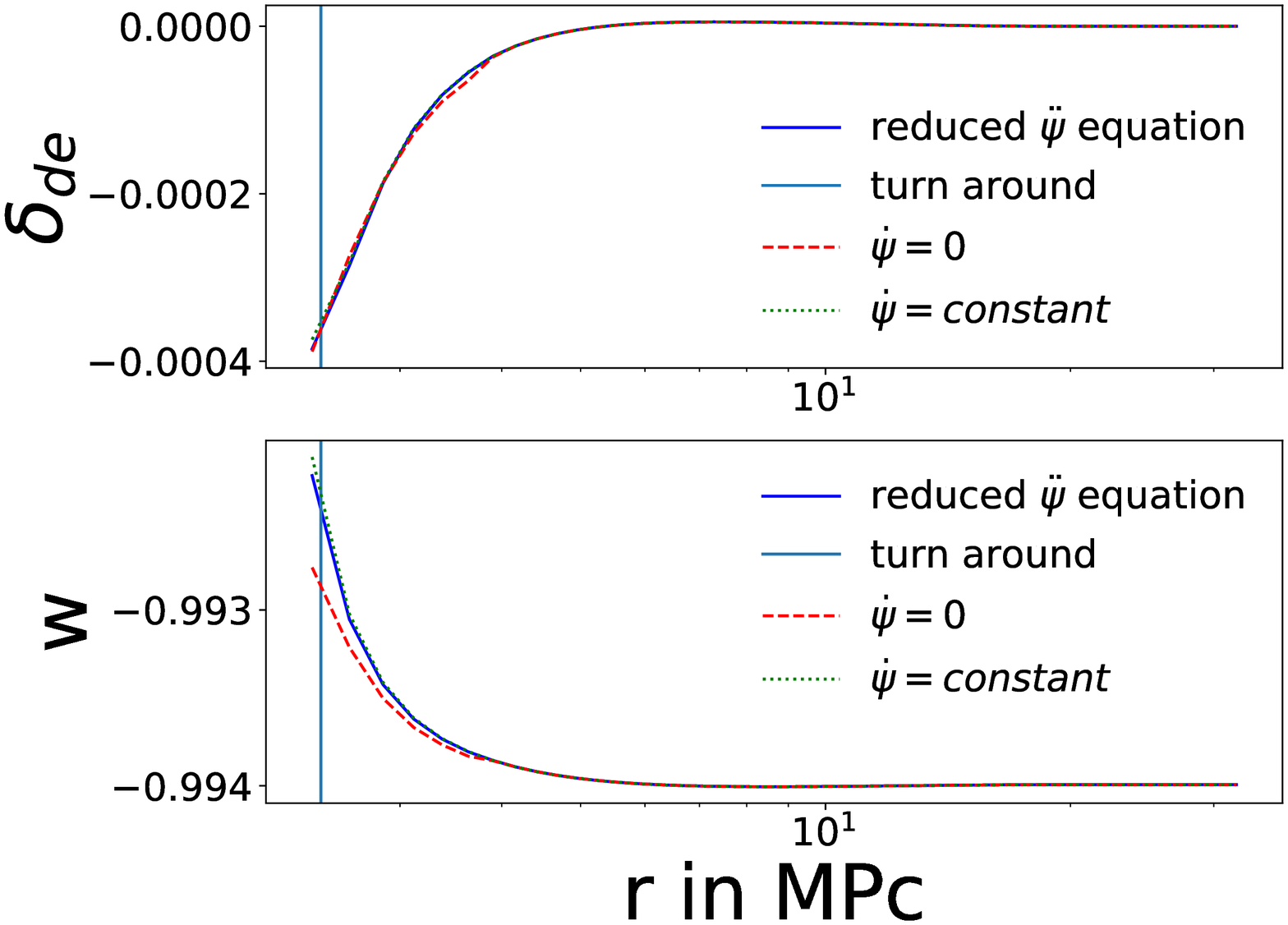}
\caption{\label{fig:0} A comparison of the three approaches for
  evolving the scalar field in the virialised region.  Here we show
  $\delta_{de}$ and $w$ as a function of $r$ at scales outside the
  virial radius at the present epoch.  The left column is for $V
  \propto \psi^2$ and the right column is for $V \propto \exp(-\psi)$
  The turn around radius is
  marked by the vertical line.  The amplitude of perturbations is
  adjusted so that the innermost shells virialise at $z \simeq 1.5$.
  Simulation OD1 was used for these plots.  
  We find that the three approaches match very well at all scales
  away from the virial radius.
  Differences between the three approaches are less than a few percent
  at all scales, and less than a percent at all scales larger than
  twice the virial radius.}
\end{figure}

A comparison of the three approaches is shown in Figure~\ref{fig:0}.
We have plotted the density contrast $\delta_{de}$ for dark energy
(top panel) and the equation of state parameter $w$ (lower panel) as a
function of scale $r$.
The two columns are for two different potentials: the left column is
for $V \propto \psi^2$ whereas the right column is for $V \propto
\exp\left[-\psi\right]$.
We have marked the turn around radius with a vertical line on these
plots. 
We find that the qualitative trend is the same for the three
approaches.
The three approaches have differences at scales close to the virial
radius, however the differences decrease rapidly beyond the turn around 
scale.
Percentage difference among three approaches outside virial
radius is less than $1\%$ at all scales.
The approach where we set $\dot\psi = 0$ deviates most from the other
two approaches and the differences are most obvious in the plot of $w$
as a function of scale $r$. 

We use the first approach where the scalar field is evolved as a test
field in the fixed background inside the virial radius in the
following discussion.

\begin{table}[tbp]
  \begin{center}
    \begin{tabular}{|c|c|c|c|c|}
      \hline
      \hline
      Simulation &$\sigma_0$&$\sigma_1$&$\alpha_0$&$z_{vir}$ \\
      \hline
      OD1 & 3 & 18 & 0.0068 & 1.5\\
      \hline
      OD2 & 3 & 18 & 0.0136 & 4.0\\
      \hline
      OD3 & 6 & 18 & 0.0068 & 1.5\\
      \hline
      UD1 & 150 & 250 & -0.0136 & -\\
      \hline
      UD2 & 20 & 200 & -0.0068 & -\\
      \hline
      UD3 & 40 & 200 & -0.0068 & -\\
      \hline
      UD4 & 20 & 200 & -0.0136 & -\\
      \hline
      UD5 & 100 & 200 & -0.0136 & -\\
      \hline
      \hline
    \end{tabular}
    \caption{\label{tab:1} Parameters used in simulations in this
      work.
      Note that for simplicity we have stated the approximate value of
      the  redshift at which the first shell virialises in the case of
      simulations with over-densities. 
      The simulations are referred to by the Simulation code in figure
      captions.} 
  \end{center}
\end{table} 

\section{Results}
\label{sec:Results}

In this section we present the results of our analysis of the system
of spherically symmetric perturbations in dark matter and dark
energy.
The complete list of simulations with the relevant parameters is given
in table~1. 
The section is divided into sub-sections where we separately study the
effect of dark energy perturbations on collapse of dark matter,
evolution of dark energy perturbations: both in the case of over
density and an under density, analysis of variations with the scale as
well as the amplitude of dark matter perturbations, and, a comparison
of the evolution of dark energy perturbations with the linear
perturbation theory. 

\begin{figure}[tbp]
\centering 
\includegraphics[width=.45\textwidth]{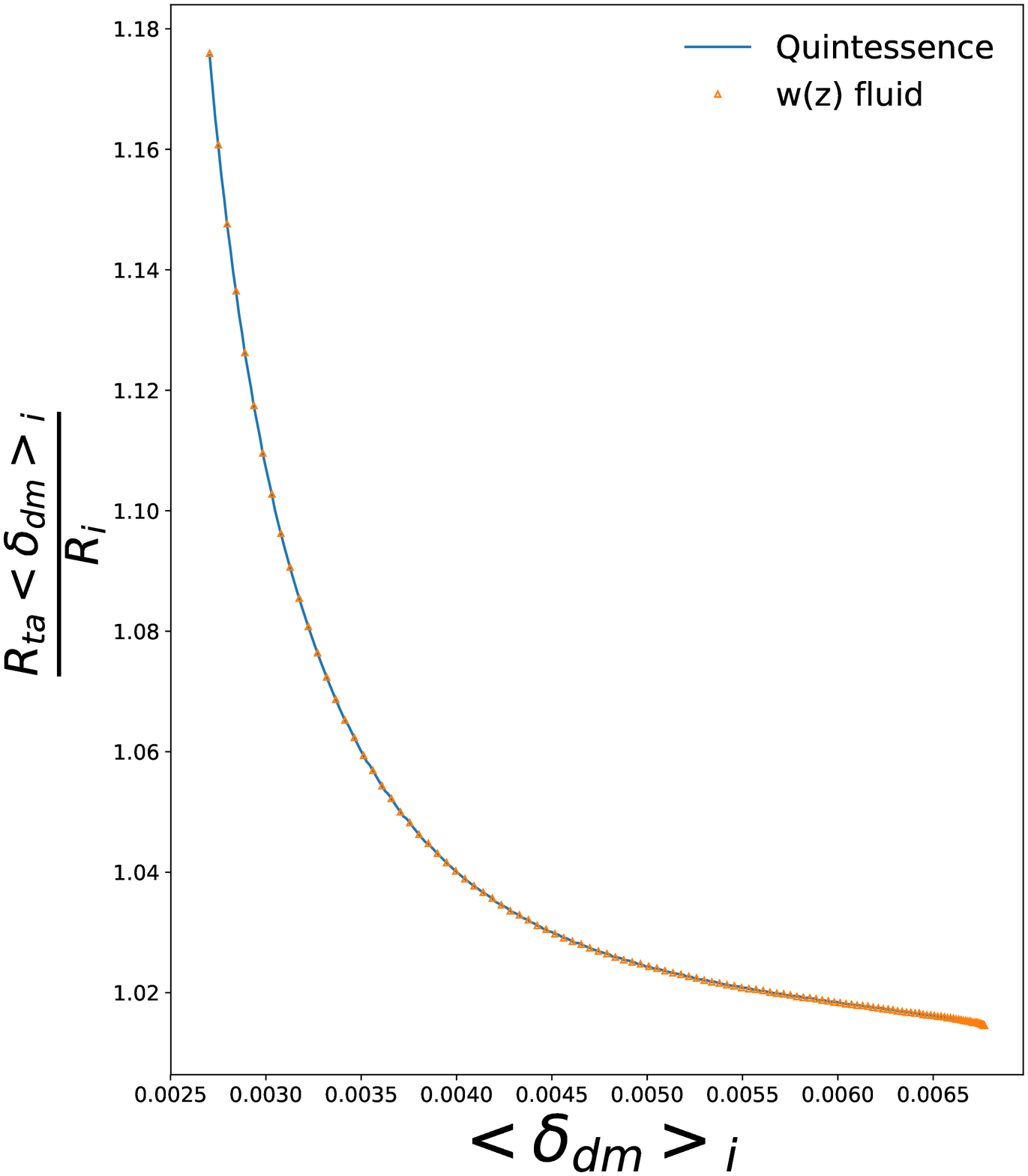}
\includegraphics[width=.45\textwidth]{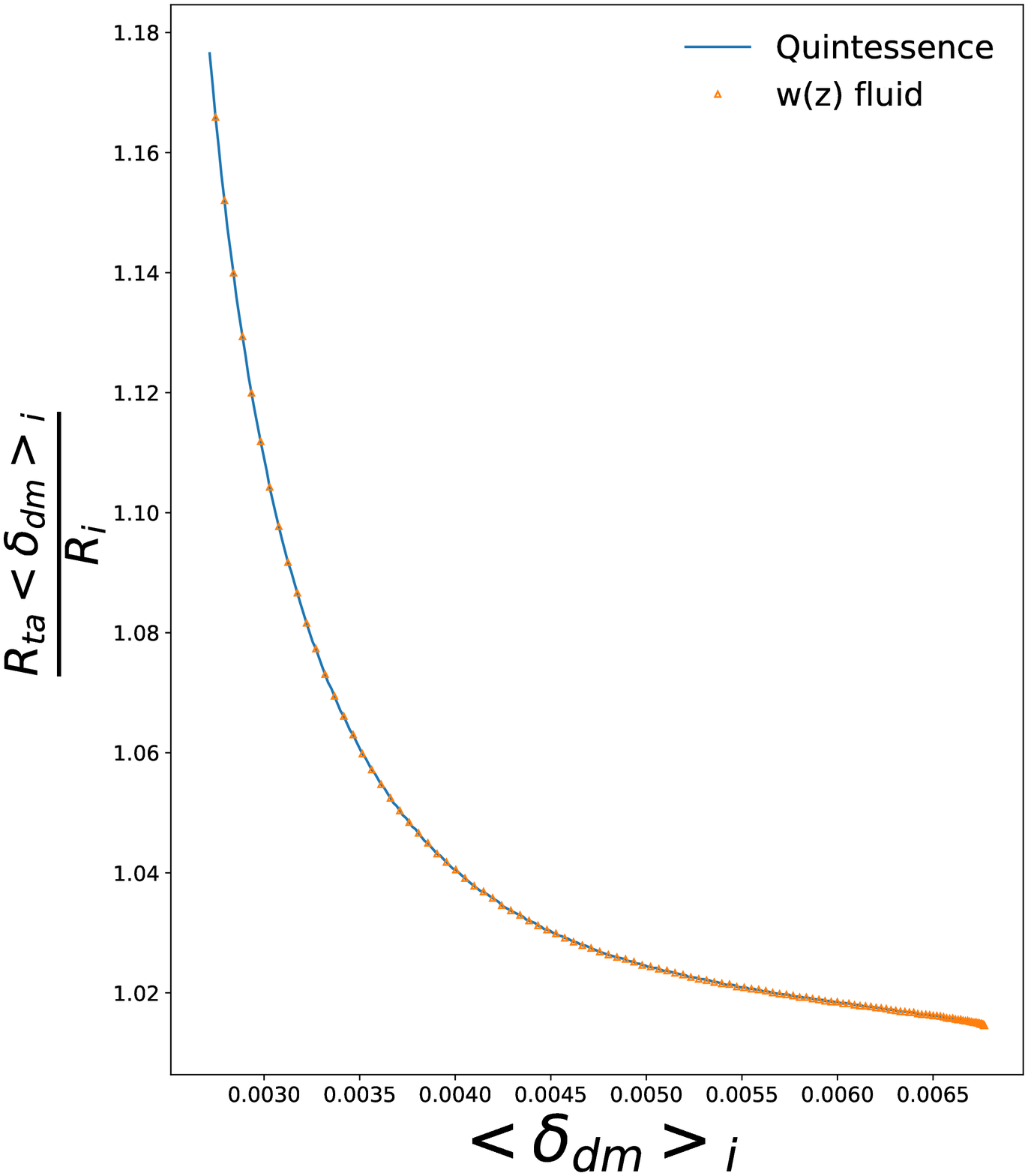}
\caption{\label{fig:3}
  We have plotted the turn around radius in the combination
  $R_{ta}\langle \delta_{dm} \rangle_i / R_i$ as a function of the
  initial density contrast.  The expected value of the combination is
  unity for the Einstein-deSitter model and we see that at large
  values of $\langle \delta_{dm} \rangle_i$ we indeed approach this
  value.  The left panel is for $V\propto \psi^2 $   while the right
  panel is for  $V\propto \exp(-\psi)$.  Simulation OD1 and the same
  initial conditions for the case without dark energy perturbations
  have been used for these plots.  We plot values from 
  our simulations with perturbations in dark energy as well as from a
  model where the dark energy does not have any perturbations.  The
  two curves match to better than $0.03\%$ at all scales, indicating
  that perturbations in dark energy do not influence collapse of dark
  matter perturbations.}
\end{figure}

\begin{figure}[tbp]
\centering 
\includegraphics[width=.45\textwidth]{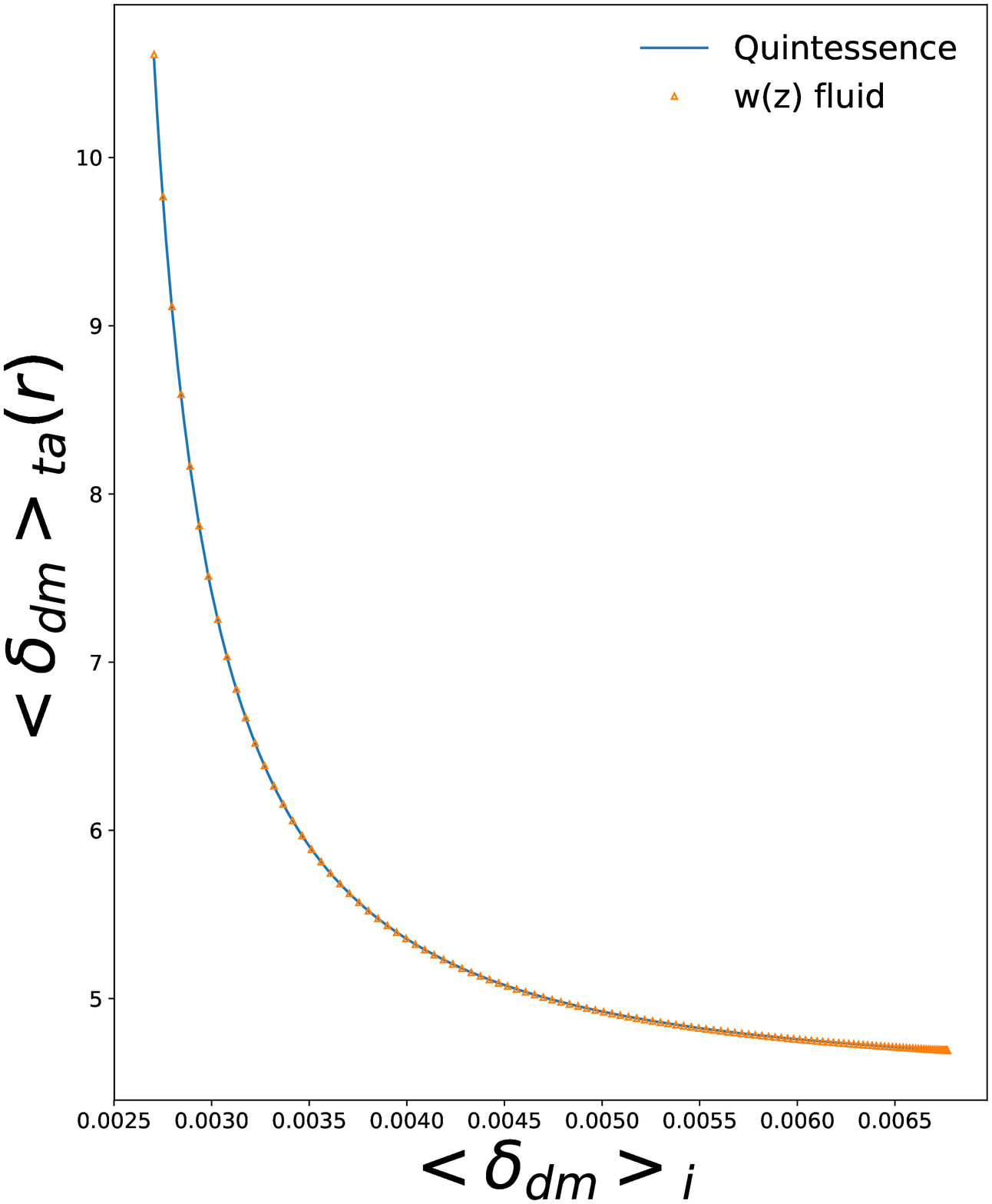}
\includegraphics[width=.45\textwidth]{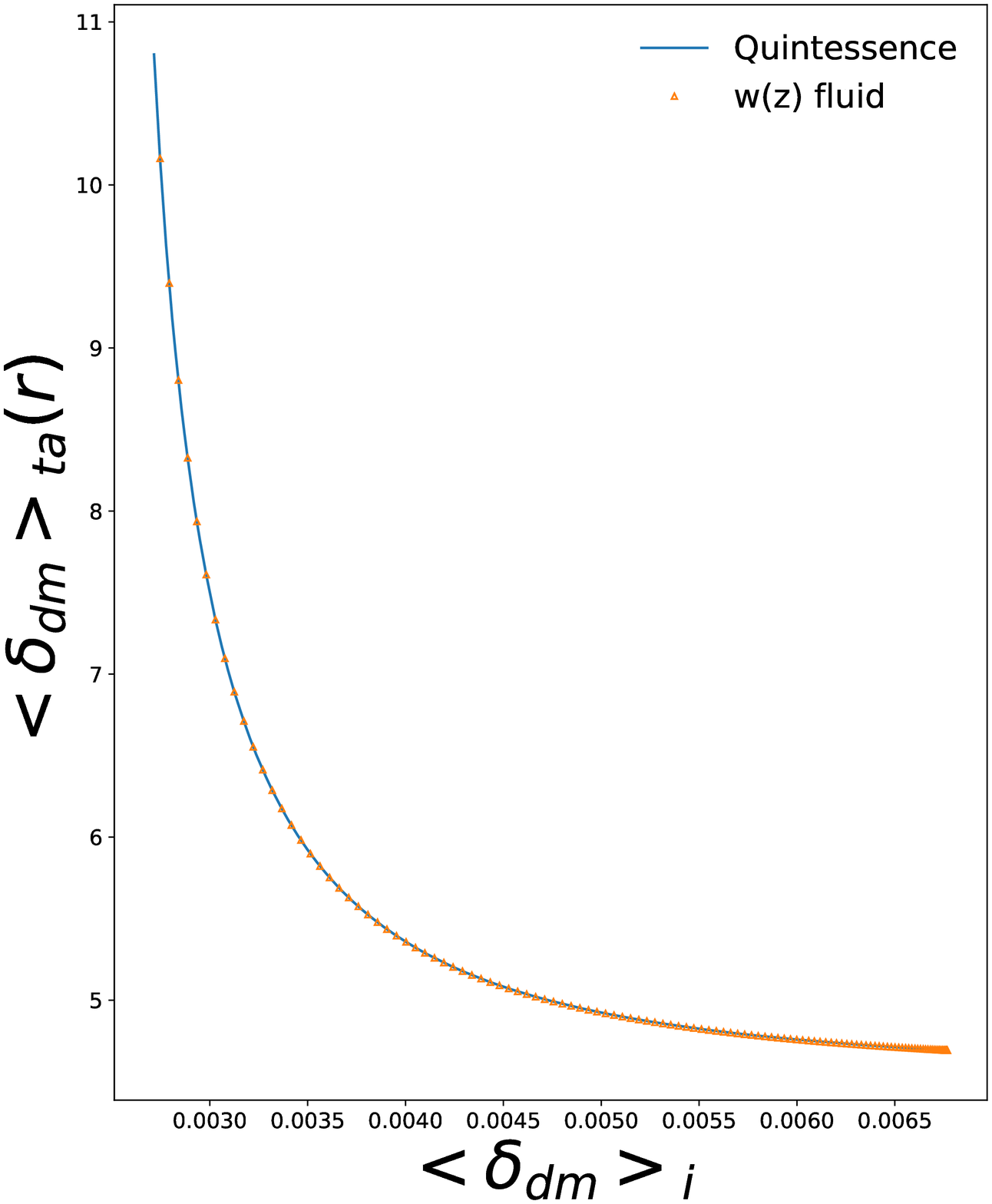}
\caption{\label{fig:4}  Density contrast at turn around is shown here
  as a function of the initial density contrast. The expected value of
  the combination is 
  $4.55$ for the Einstein-deSitter model and we see that at large
  values of $\langle \delta_{dm} \rangle_i$ we indeed approach this
  value.  The left panel is for $V\propto \psi^2 $   while the right
  panel is for  $V\propto \exp(-\psi)$.  We plot values from 
  our simulations with perturbations in dark energy as well as from a
  model where the dark energy does not have any perturbations.
  Initial dark matter perturbations here correspond to simulation
  OD1.  The
  two curves match to better than $0.06\%$ at all scales, indicating
  that perturbations in dark energy do not influence collapse of dark
  matter perturbations.  
}
\end{figure}

\begin{figure}[tbp]
\centering 
\includegraphics[width=.45\textwidth]{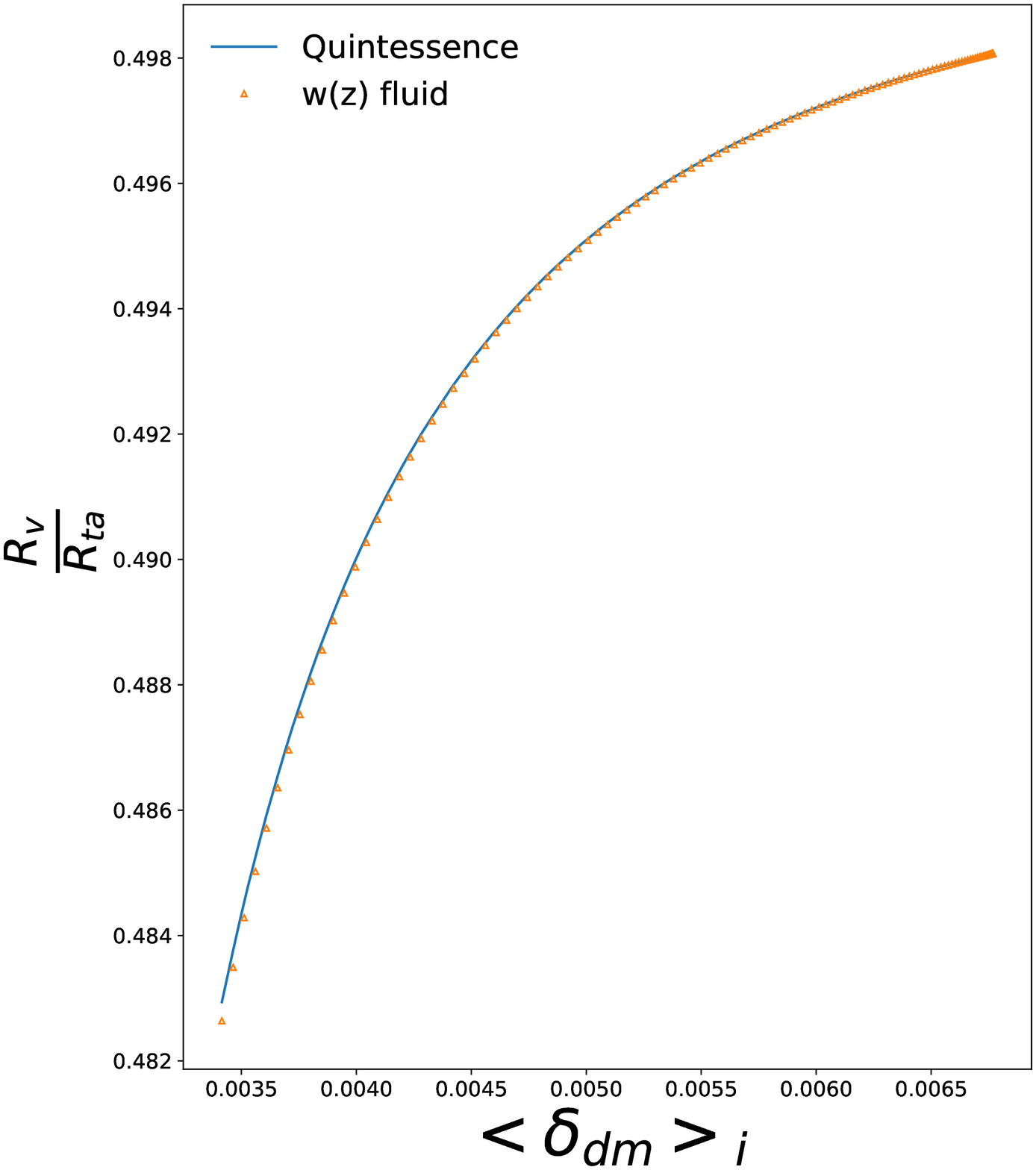}
\includegraphics[width=.45\textwidth]{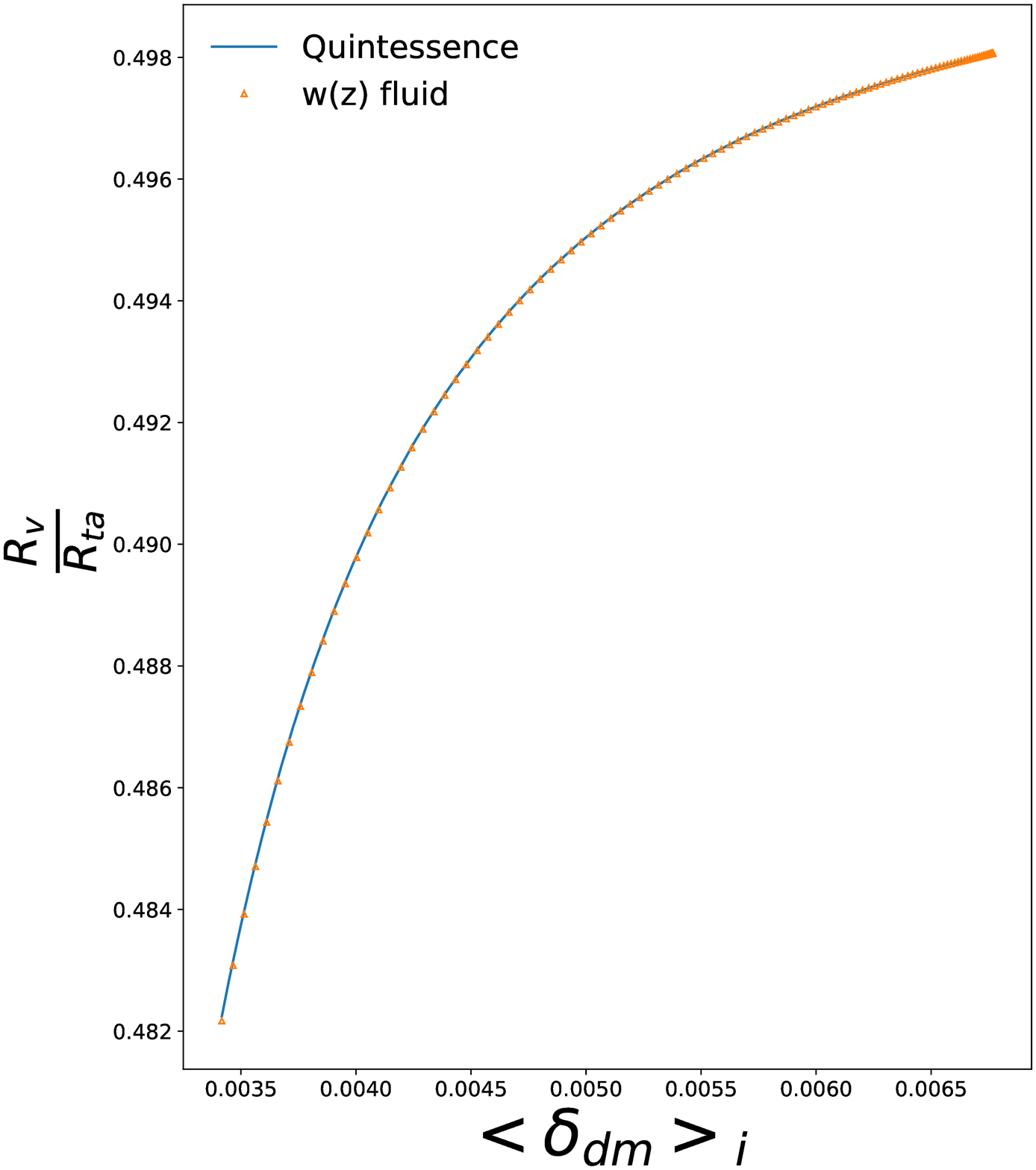}
\caption{\label{fig:5} Ratio of virial radius to turn around
  radius is shown here as a function of the initial density contrast
  in dark matter.  The expected value of
  the combination is 
  $0.5$ for the Einstein-deSitter model and we see that at large
  values of $\langle \delta_{dm} \rangle_i$ we indeed approach this
  value.  A value lower than $0.5$ signifies that dark energy does not
  cluster significantly \cite{2005JCAP...07..003M}. 
  The left panel is for $V\propto \psi^2 $   while the right
  panel is for  $V\propto \exp(-\psi)$.  We plot values from 
  our simulations (OD1) with perturbations in dark energy as well as
  from a model where the dark energy does not have any perturbations.
  The two curves match to better than $0.01\%$ at all scales,
  indicating that perturbations in dark energy do not influence
  collapse of dark matter perturbations.}
\end{figure}

\begin{figure}[tbp]
\centering 
\includegraphics[width=.45\textwidth]{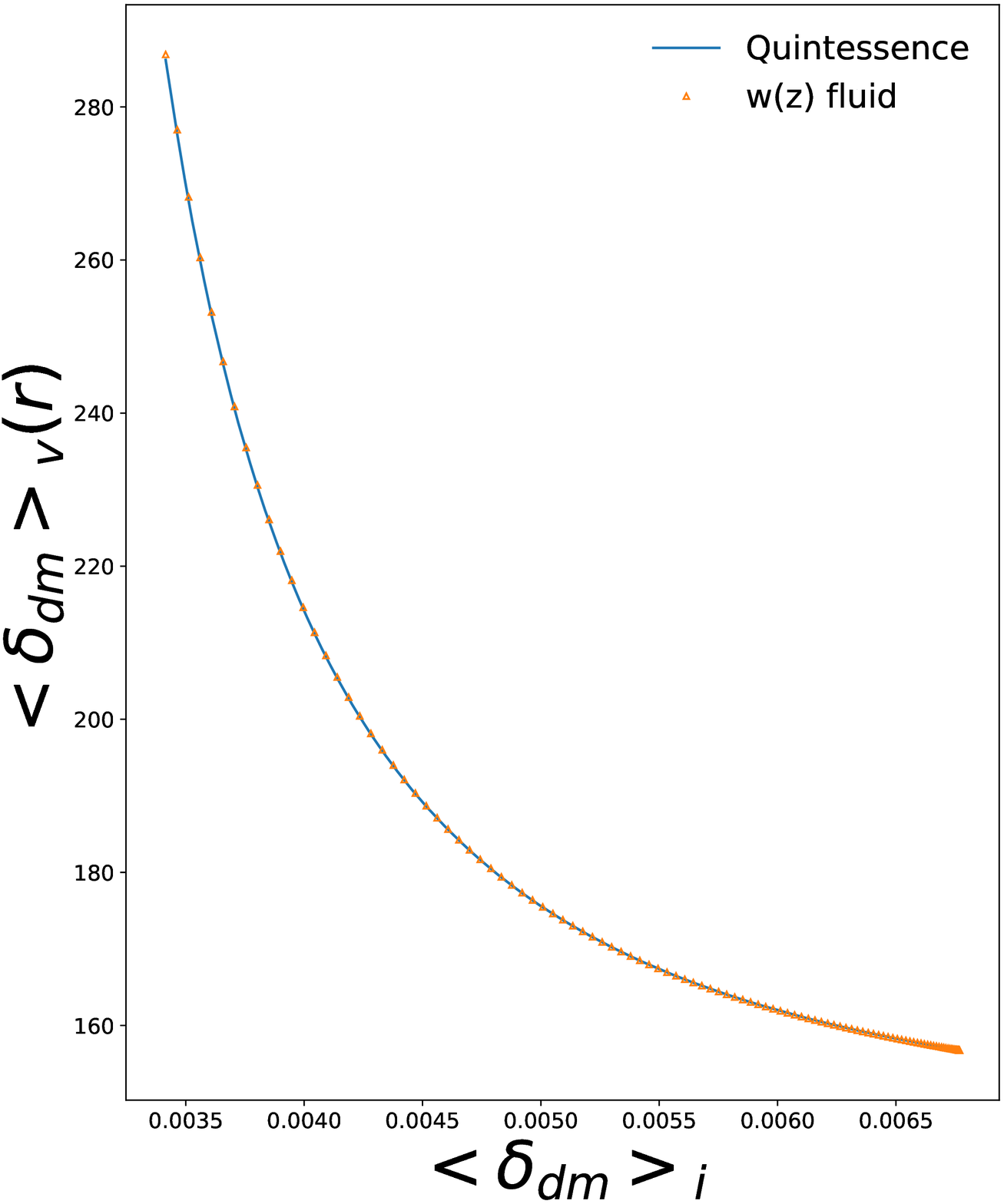}
\includegraphics[width=.45\textwidth]{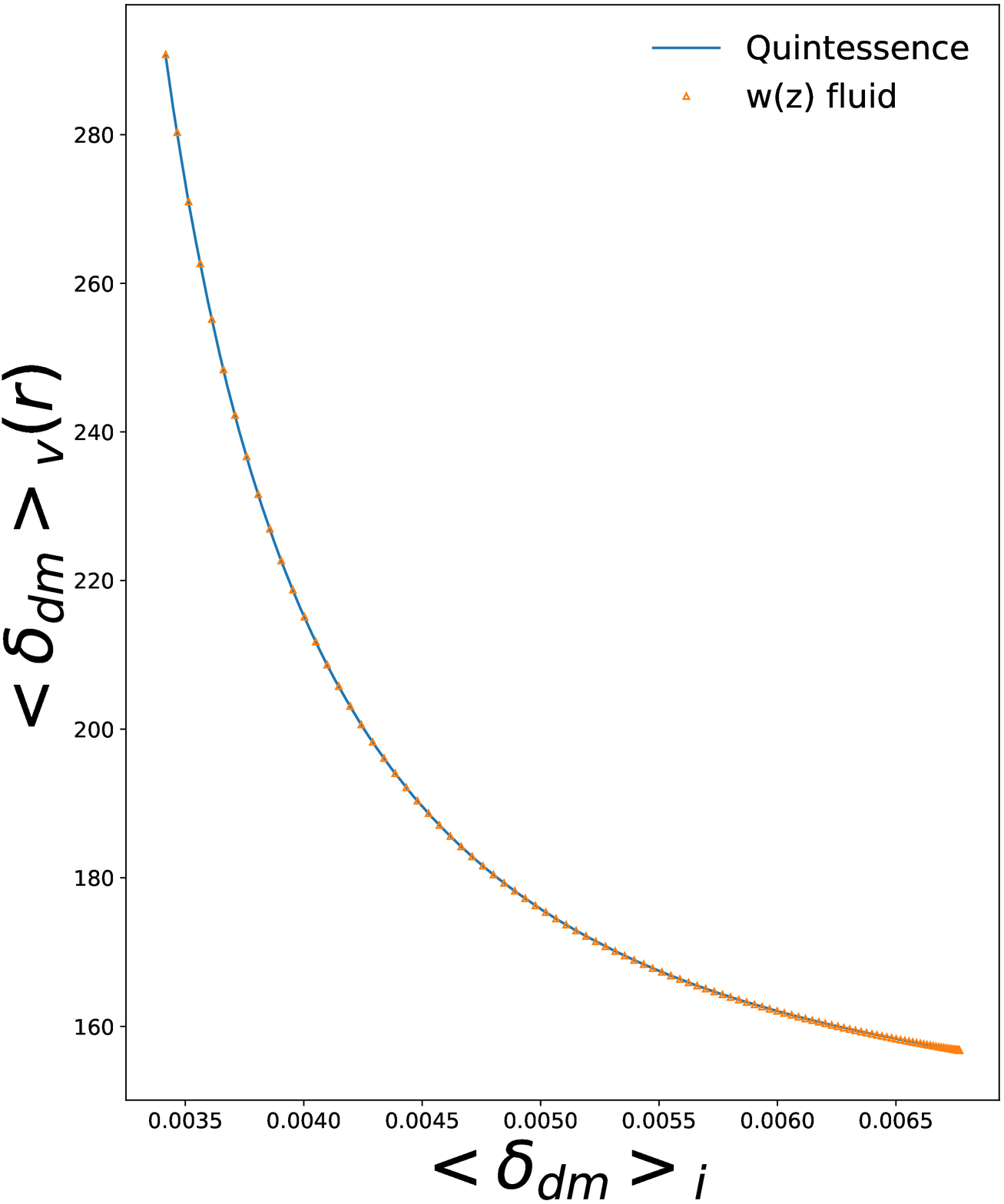}
\caption{\label{fig:6}
  Density contrast at virialisation as a function of the initial
  density contrast in dark matter.  The expected value in our approach
  for Einstein-deSitter universe is $145$ (see text).  We see that the
  curve is tending towards that value at large initial density
  contrast.  The left panel is for $V\propto \psi^2 $   while the right
  panel is for  $V\propto \exp(-\psi)$.  We plot values from 
  our simulations (OD1) with perturbations in dark energy as well as
  from a model where the dark energy does not have any perturbations.
  The two curves match to better than $0.3\%$ at all scales,
  indicating that perturbations in dark energy do not influence
  collapse of dark matter perturbations.  
}
\end{figure}

\subsection{Dark matter perturbations}
\label{subsec:dmp}

In order to study the effects of DE perturbations of dark matter, we
ran, besides simulations mentioned above, a dark energy model
simulation by taking $w(z)$ from the same background and then
implementing a non-clustering fluid with same numerical evolution of
$w$ to act as DE with following dynamics:
\begin{equation}
\frac{d{\rho_{de}}}{dt}=-3\frac{\dot{a}}{a}(1+w)\rho_{de}
\end{equation} 

Results of a comparative study of this fluid model with full-fledged
spherical collapse in quintessence are presented here. 
We show the comparison for various quantities related to turn around
and virialisation.
We choose to plot these as a function of the initial density contrast
$\delta_i$.
The choice is motivated by the emergence of a critical value for
density contrast required for collapse in the case of the cosmological
constant.
We find that just like the cosmological constant model for dark
energy, there is a critical value that emerges in the dynamical dark
energy models.
Perturbations with a lower initial density contrast do not enter a
collapsing phase.
Further, we find that various quantities of interest approach the
values obtained in the Einstein-deSitter model as $\delta_i$ becomes
much larger than the critical value.
On the other hand, as we approach the critical initial density
contrast from above, dark energy becomes more and more important, and
hence it takes longer to begin collapse.
Thus the universe expands by a significantly larger amount by the time
such perturbations reach turn around or virialisation and hence the
average density of matter in the universe is much lower.  

Figure~\ref{fig:3} shows the turn around radius as a function of
$\delta_i$.
Instead of the turn around radius, we choose to plot the combination
$R_{ta} \langle \delta_i \rangle / R_i$.
Here $R_i$ is the initial radius of the shell and $\langle \delta_i
\rangle$ is the average density contrast inside this shell at the
initial time.
This combination is unity for spherical collapse in the
Einstein-deSitter model.
The top left panel is for the $\psi^2$ potential whereas the top right
panel is for the exponential potential.
Curve for the model with dark energy perturbations and points for the 
corresponding model without dark energy perturbations are plotted in
the same panels.
The difference between the two cases is too small to be seen from
these panels. 
In both models, and for the cases with and without dark energy
perturbations, the qualitative trend is the same: the turn around
radius is larger for smaller $\delta_i$.
At large $\delta_i$, we approach the turn around radius approaches the
expected value in the Einstein-deSitter model.
We find that the percentage difference is well below one percent for
the turn around radius.

Figure~\ref{fig:4} shows the turn around density contrast for
different shells in the same format as Figure~\ref{fig:3}.
We find that the density contrast at turn around for shells with large
$\delta_i$ approaches the expected value for the Einstein-deSitter
model.
As we approach lower $\delta_i$, we find that the density contrast at
turn around increases rapidly.
This is largely because it takes longer to reach turn around for shells
with a smaller initial density contrast, and in this time the density
of matter in the universe decreases significantly, leading to a larger
density contrast within the perturbation. 
In this case too, the difference between the model with dark energy
clustering and without dark energy clustering is smaller than a
percent at all scales for the two potentials studied here. 

Figure~\ref{fig:5} and Figure~\ref{fig:6} show the virial radius (in
units of the turn around radius) and the density contrast at the time
of virialisation, respectively.
We find that the two quantities approach the values expected for the
Einstein-deSitter model at large $\delta_i$.
For shells with smaller $\delta_i$, the virial radius is less than
half the turn around radius with the ratio decreasing as we get to
shells with a smaller initial $\delta_i$.
The density contrast at virialisation increases rapidly for smaller
initial $\delta_i$, whereas for larger $\delta_i$, we get the 
value expected  in the Einstein-deSitter model ($145$).

\subsection{Dark Energy Perturbations}
\label{subsec:DEP}

Here we study the evolution of dark energy perturbations in spherical
collapse.
As stated above, the dark energy component does not have any initial
perturbations and therefore perturbations evolve in response to the
density contrast in dark matter. 
The discussion is divided into two segments, one each for an
over density in matter, and an under density in matter.

\begin{figure}[tbp]
\centering 
\includegraphics[width=.45\textwidth]{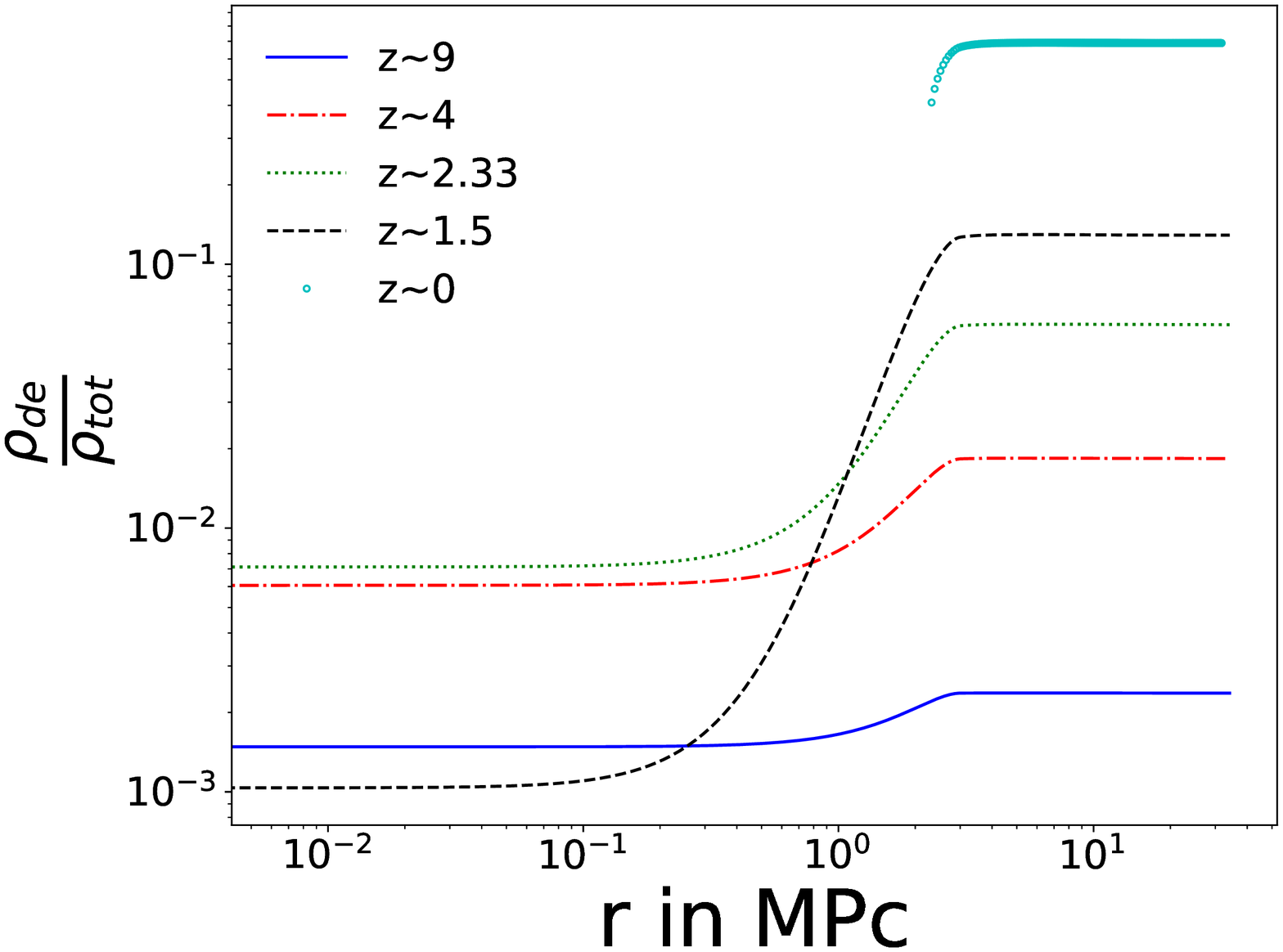}
\includegraphics[width=.45\textwidth]{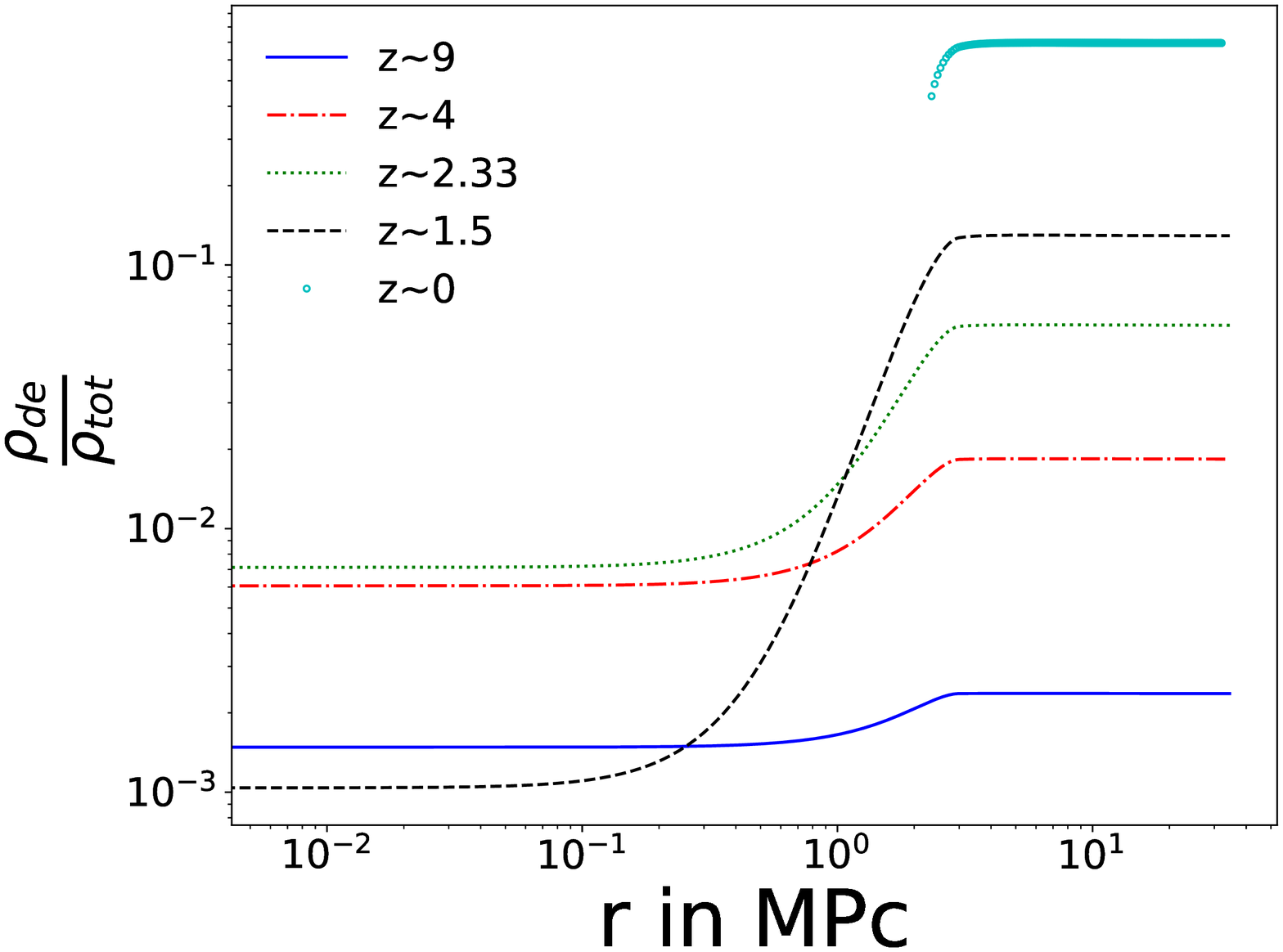}
\caption{\label{fig:8} Contribution of Dark Energy to total energy
  density for over dense case (simulation OD1).  We see that at large
  $r$, the relative 
  contribution of dark energy increases monotonically.  However,
  within the over dense region the contribution of dark energy reaches
  a maximum of $\sim 0.007$ and then drops to lower values.  The left
  panel is for $V\propto \psi^2 $ while the right panel is for
  $V\propto \exp(-\psi) $.  We have shown curves outside the virial
  radius while omitting the values inside the virial radius as we do
  not have a self-consistent evolution inside the virialised halo.
  This omission of data within the virial radius impacts only one of
  the curves shown here.}
\end{figure}

\begin{figure}[tbp]
\centering 
\includegraphics[width=.45\textwidth]{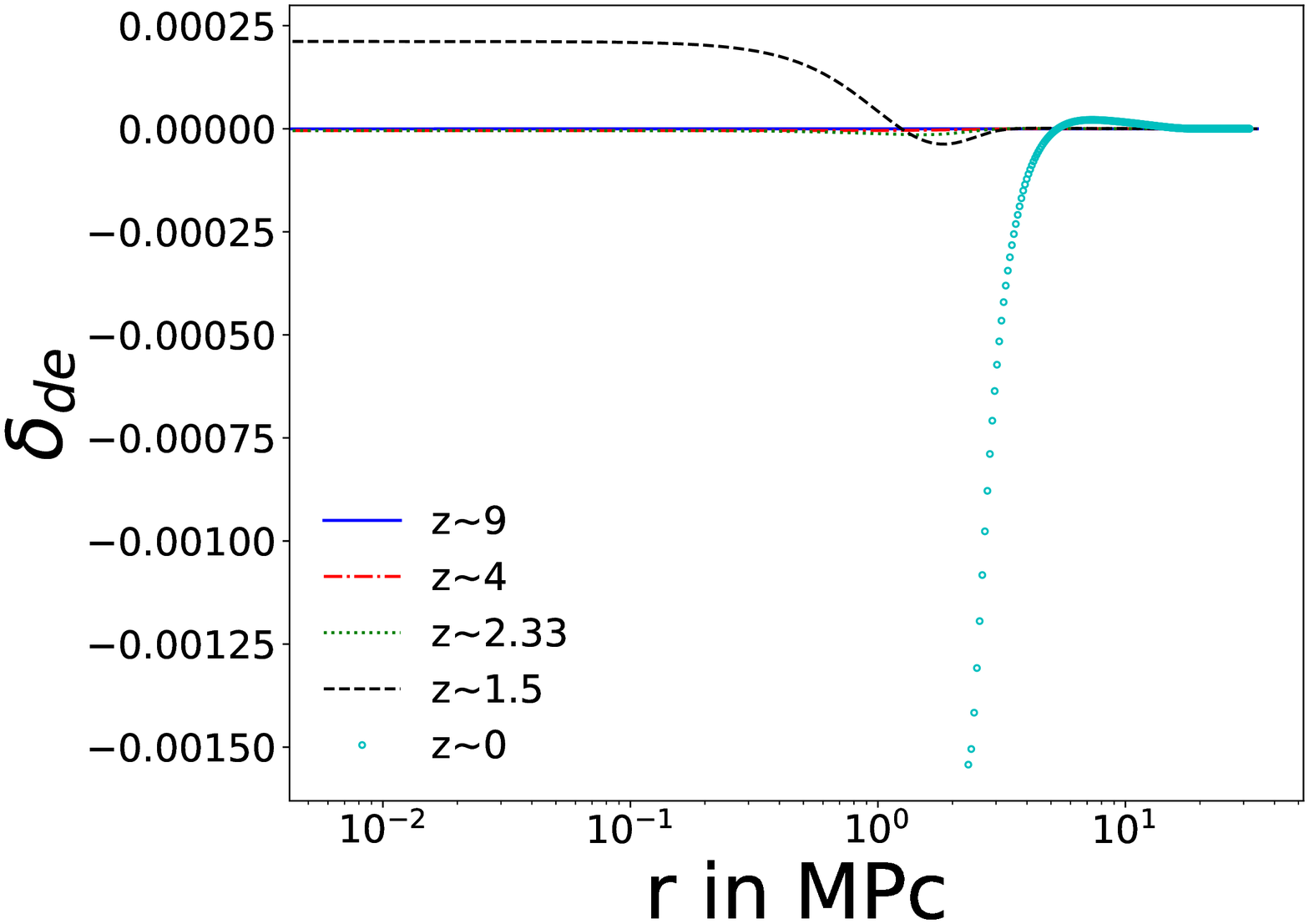}
\includegraphics[width=.45\textwidth]{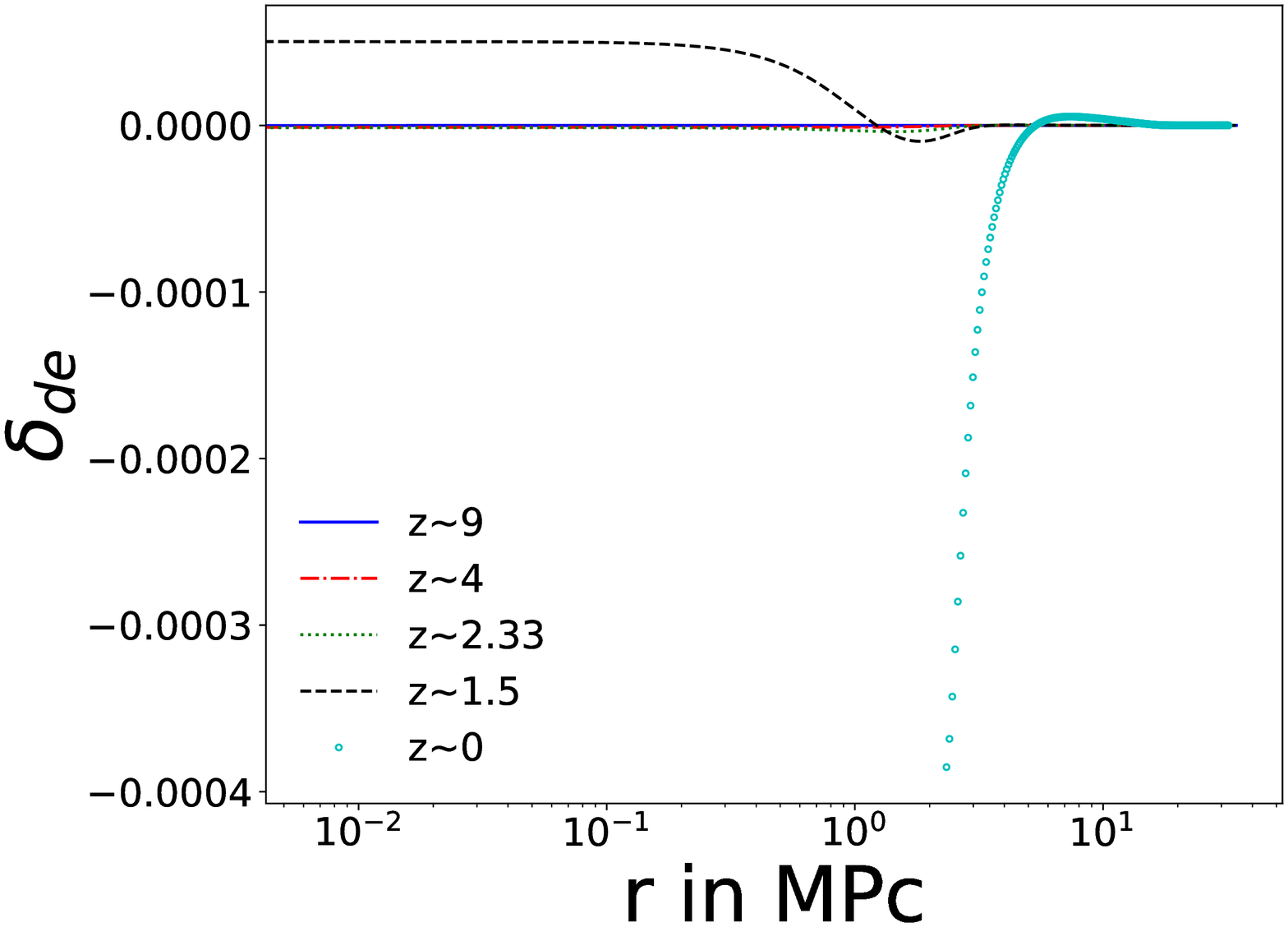}
\caption{\label{fig:9}
  Density contrast for dark energy as a function of scale at different
  epochs.  We see that the amplitude of perturbations in dark energy
  remains small at all scales and at all times.  We have plotted
  values only at scales outside the virial radius for simulation OD1.
  The left panel is for $V\propto \psi^2 $ while the right panel is
  for $V\propto \exp(-\psi) $.}
\end{figure}

\begin{figure}[tbp]
\centering 
\includegraphics[width=.45\textwidth]{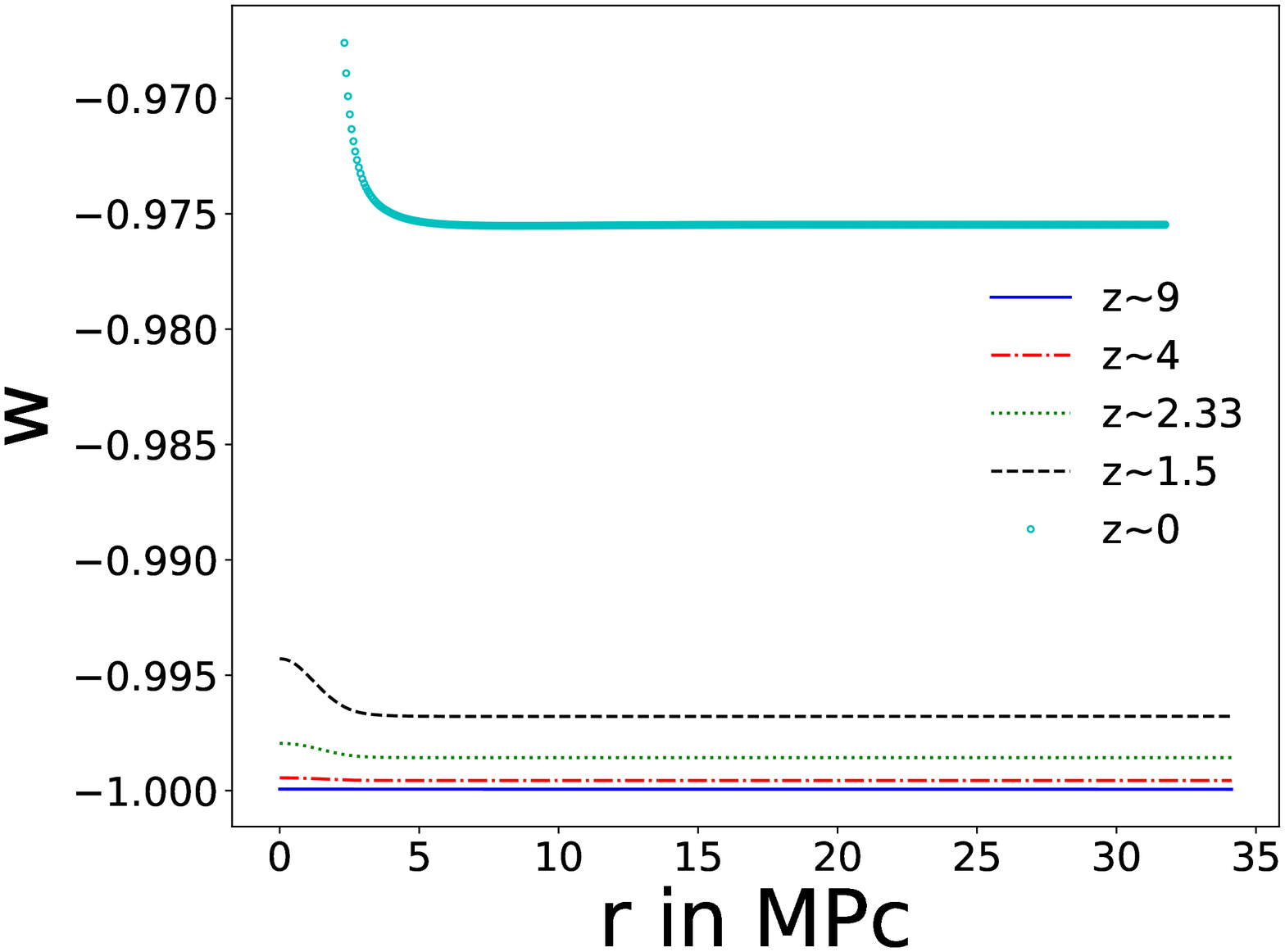}
\includegraphics[width=.45\textwidth]{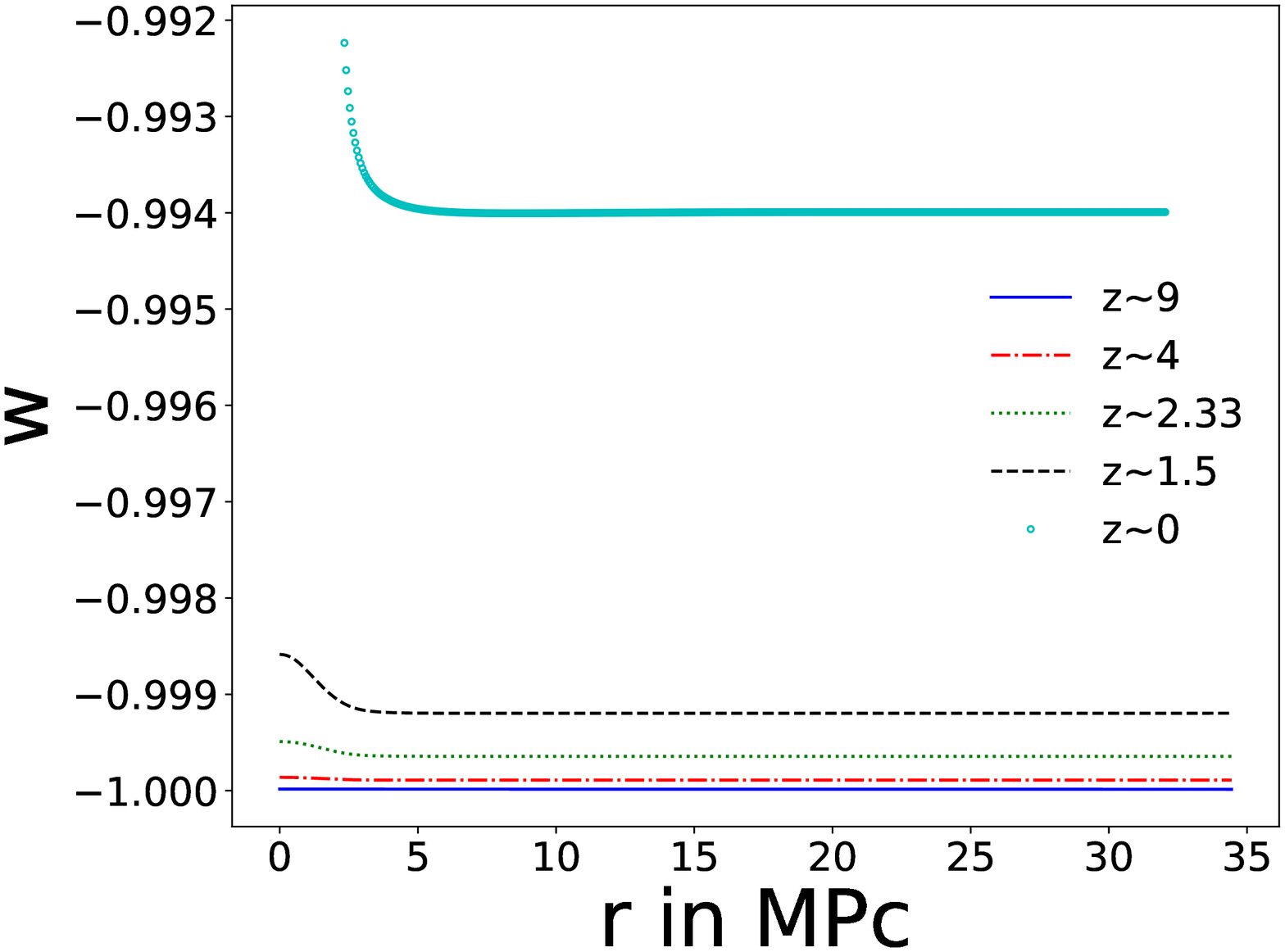}
\caption{\label{fig:10}  The equation of state parameter for dark
  energy as a function of scale at different epochs from simulation
  OD1.  We see that the 
  variation in $w$ with scale is fairly significant, particularly at
  late times.  The lower row of plots shows the variation of $w$ with
  respect to the value in the background model, or the asymptotic
  value at large scales. 
  We have plotted
  values only at scales outside the virial radius.  The left
  panel is for $V\propto \psi^2 $ while the right panel is for
  $V\propto \exp(-\psi) $.}
\end{figure}

\subsubsection{Over dense Profile}
\label{subsubsec:DEP:OD}

In a region with an initial over density in matter, gravitational
instability ensures that the density of dark matter in the region
increases monotonically when compared with the average density of
matter in the universe. 
If the initial density contrast is sufficiently high, we find that
gravitational instability leads to local collapse and a sharp increase
in the density of matter within the collapsed region.
It is important to assess the evolution of dark energy density in the
region.
We show the relative contribution of energy density of dark energy as
compared to dark matter drops significantly in the region where dark
matter collapses.
We show this for a model with $\sigma_0=3$ Mpc,
  $\sigma_1=18$ Mpc and the redshift of virialisation $z\sim 1.5$.
This is shown in Figure~\ref{fig:8} 
Each curve refers to a specific epoch as marked in the legend.
We find that at very large scales dark energy becomes more and more
dominant with time, as is expected for the background model that is
dominated by dark energy at present.
However, within the collapsed region, the relative role of dark energy
diminishes strongly at late times. 
We see that even before virialisation, the energy density of dark
energy drops to less than a few percent of its background value near
the centre of the perturbation.
Thus in terms of the local contribution to the energy budget, dark
energy plays a negligible role inside the perturbation.

We plot the density contrast for dark energy as a function of scale
for the same model used above.
We find that the density contrast for dark energy grows in response to
the dark matter perturbation, however its amplitude remains small as
compared to unity through the non-linear evolution of dark matter
perturbations. 
Thus we do not expect any significant impact of dark energy density
contrast and its variations on observables at small scales.

A surprising feature that may have implications for observations and
hence work as a diagnostic for dynamical dark energy models is the
spatial variation in the equation of state parameter $w$.
We already know from background evolution and our choice of initial
conditions that $w=-1$ at early times and it increases slowly with
time. 
We show variation of $w$ as a function of $r$ in Figure~\ref{fig:10}.
This is shown for four epochs leading up to the epoch of
virialisation.
We find that $w$ increases more rapidly in regions around the
collapsing dark matter perturbations.

\begin{figure}[tbp]
\centering 
\includegraphics[width=.45\textwidth]{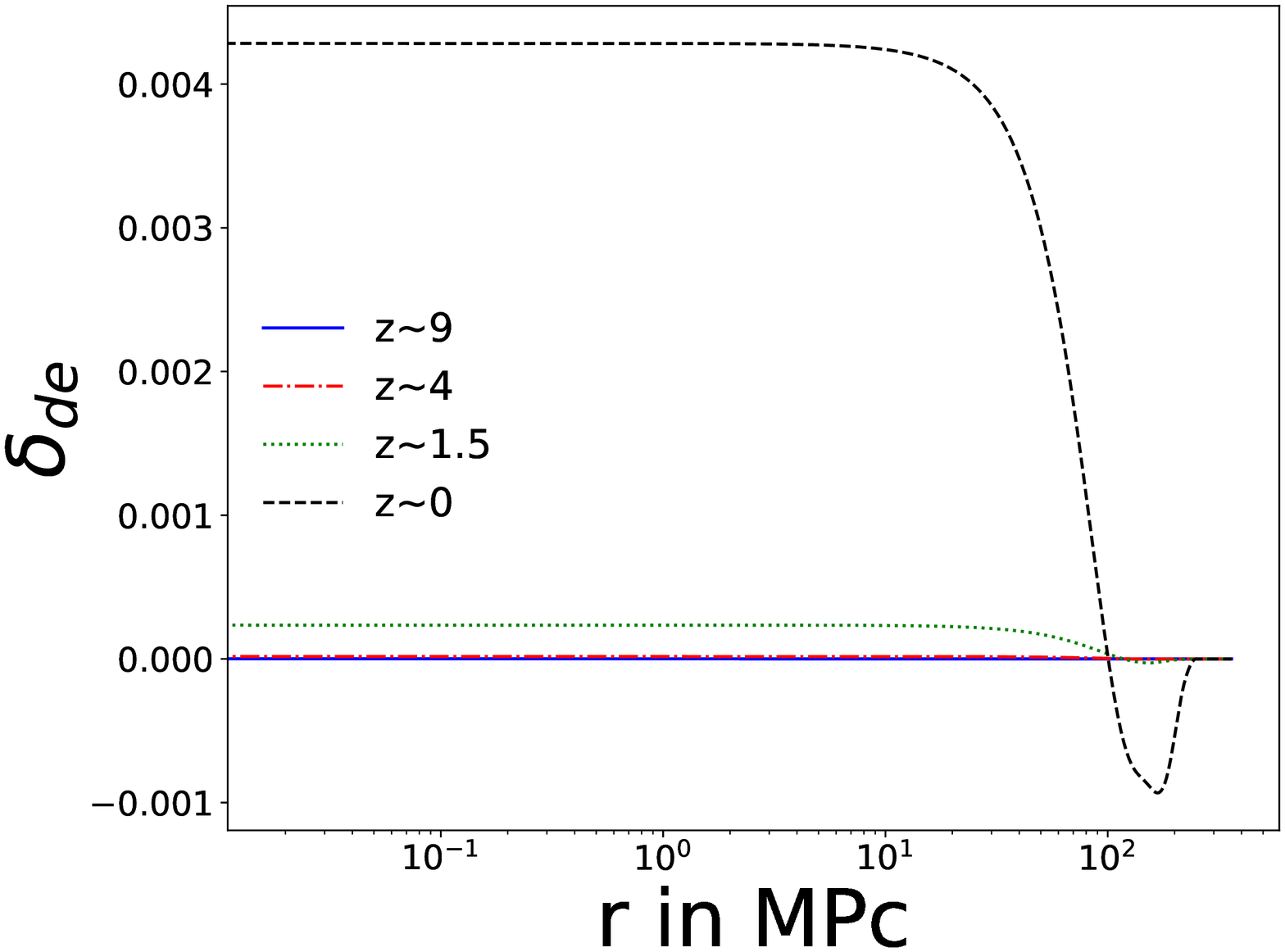}
\includegraphics[width=.45\textwidth]{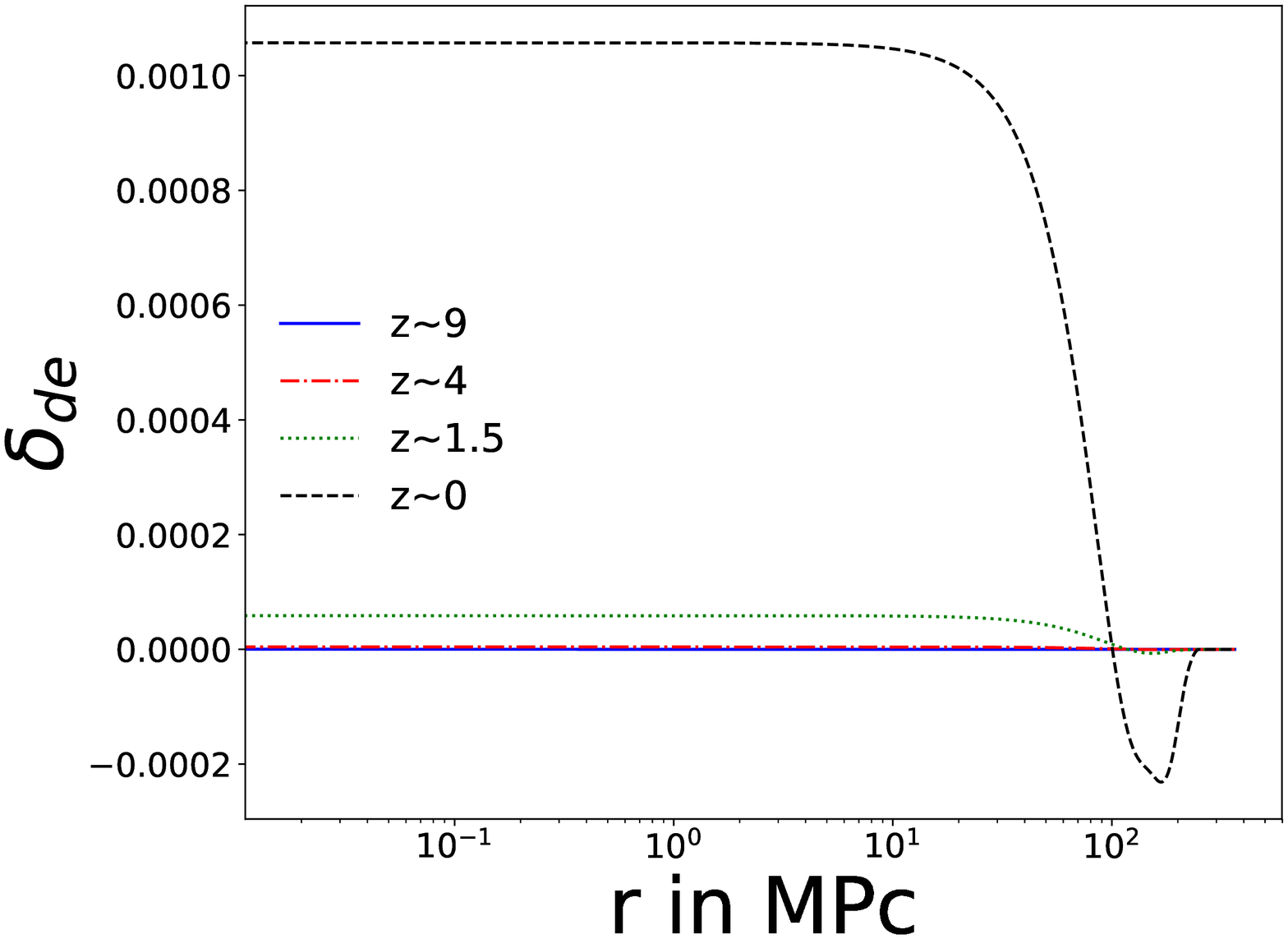}
\caption{\label{fig:12} Density contrast for dark energy as a function
  of scale $r$ for a void, i.e., a matter under-density from simulation
  UD1.  This is
  plotted at multiple epochs.  We find that dark energy perturbations
  grow but the amplitude remains small in absolute terms.   The left
  panel is for $V\propto \psi^2 $ while the right panel is for
  $V\propto \exp(-\psi) $. }
\end{figure}

\begin{figure}[tbp]
\centering 
\includegraphics[width=.45\textwidth]{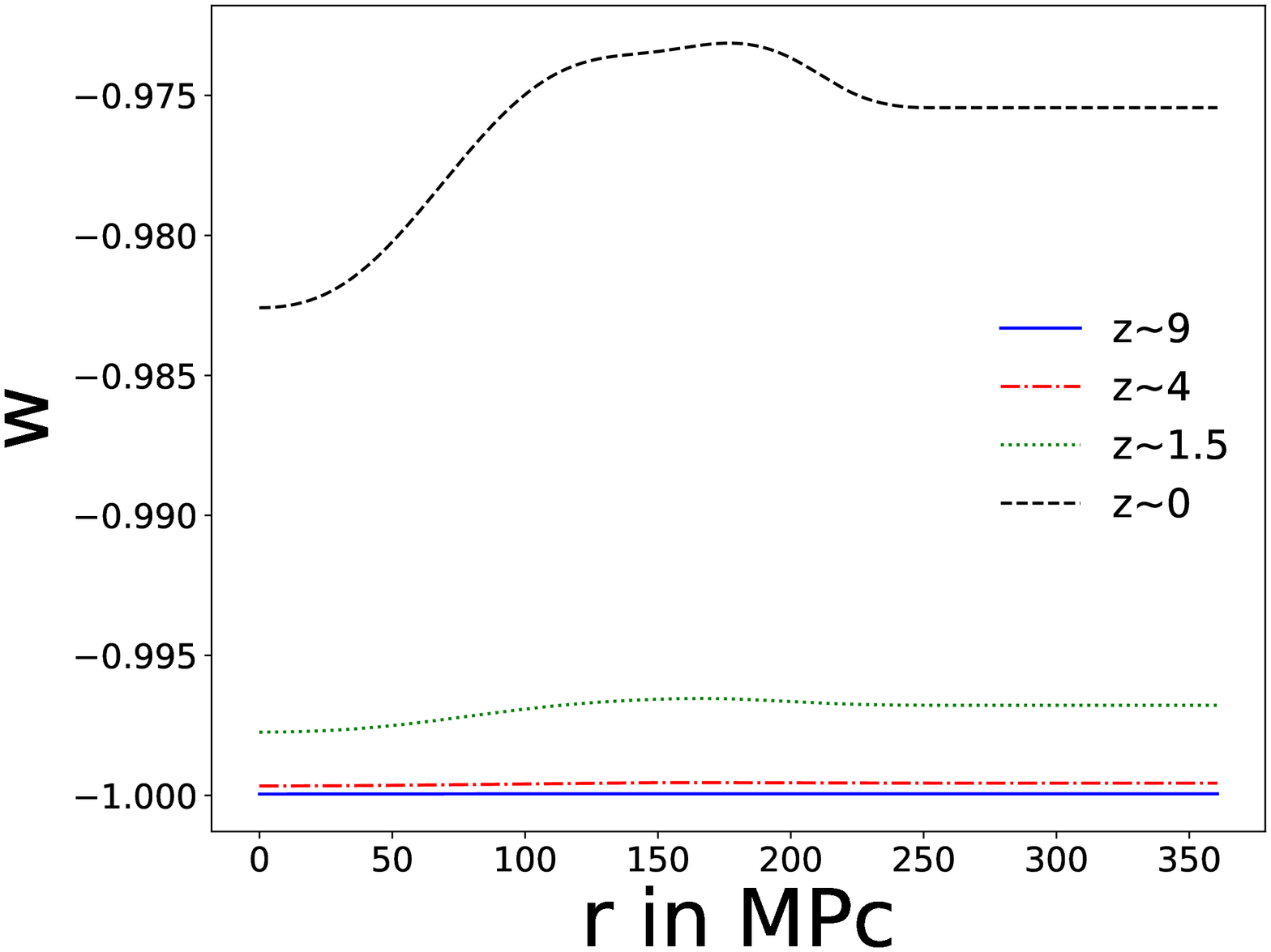}
\includegraphics[width=.45\textwidth]{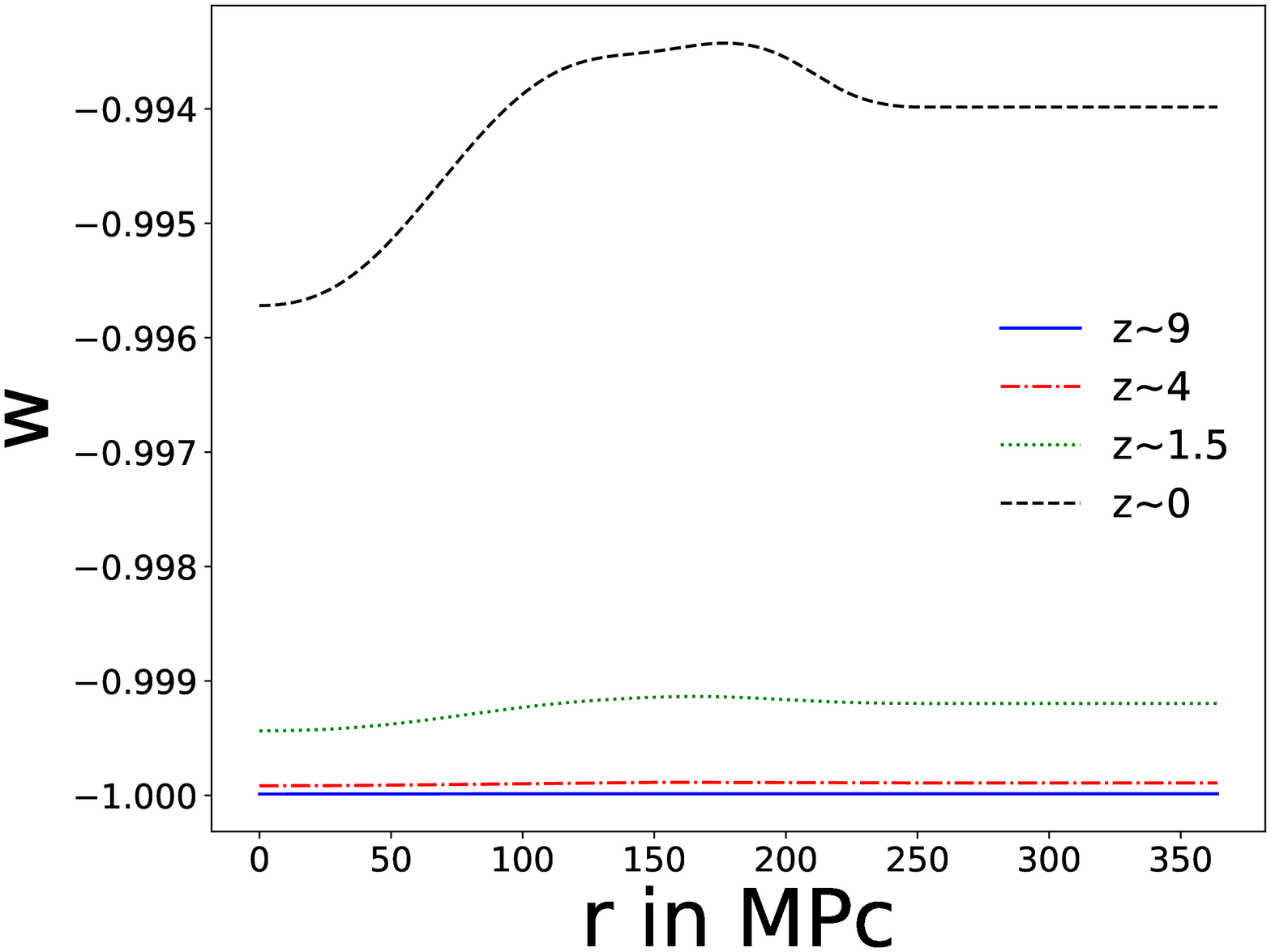}
\caption{\label{fig:13} Equation of state parameter $w$ as a function
  of scale $r$ for a void, i.e., a matter under-density for simulation
  UD1.  This is
  plotted at multiple epochs.  We find that $w$ inside the void is
  smaller than at large scales.  The left
  panel is for $V\propto \psi^2 $ while the right panel is for
  $V\propto \exp(-\psi) $.}
\end{figure}

\subsubsection{Under dense Profile}
\label{subsubsec:DEP:UD}

So far we have discussed the evolution of matter over densities.
We now turn our attention to the evolution of under densities, or
voids.
The large scale of voids coupled with the fact that the magnitude of
the spatial variation of $w$ is larger for perturbations at large
scales makes these a potential test bed for observing the effects of
dynamical dark energy. 

We show results for a model with $\sigma_0=150$ and $\sigma_1=250$,
thus the characteristic length scale of the perturbation is $250$.
We find that the dark energy contributes a very significant fraction
to the total energy budget mainly due to depletion of matter.
This becomes clear in figure~\ref{fig:12} that shows the density
contrast in dark energy as a function of scale $r$.
We find that the amplitude of density contrast is very small compared
to unity at all scales and at all times. 

We have plotted the variation of $w$, the equation of state parameter,
as a function of scale at different epochs in figure~\ref{fig:13}.
We find that the increase in $w$ with time slows down in under dense
regions.
This is mainly due to the faster than average expansion rate in the
voids.
We find that the differential in $w$ is larger for larger voids.
The variation with the initial density contrast for matter is less
significant, but a larger initial under density leads to a larger
differential in $w$.

Voids may be the optimal sites for testing changes in $w$.
This is primarily because dark energy dominates in terms of the
overall energy budget.

\begin{figure}[tbp]
\centering 
\includegraphics[width=.45\textwidth]{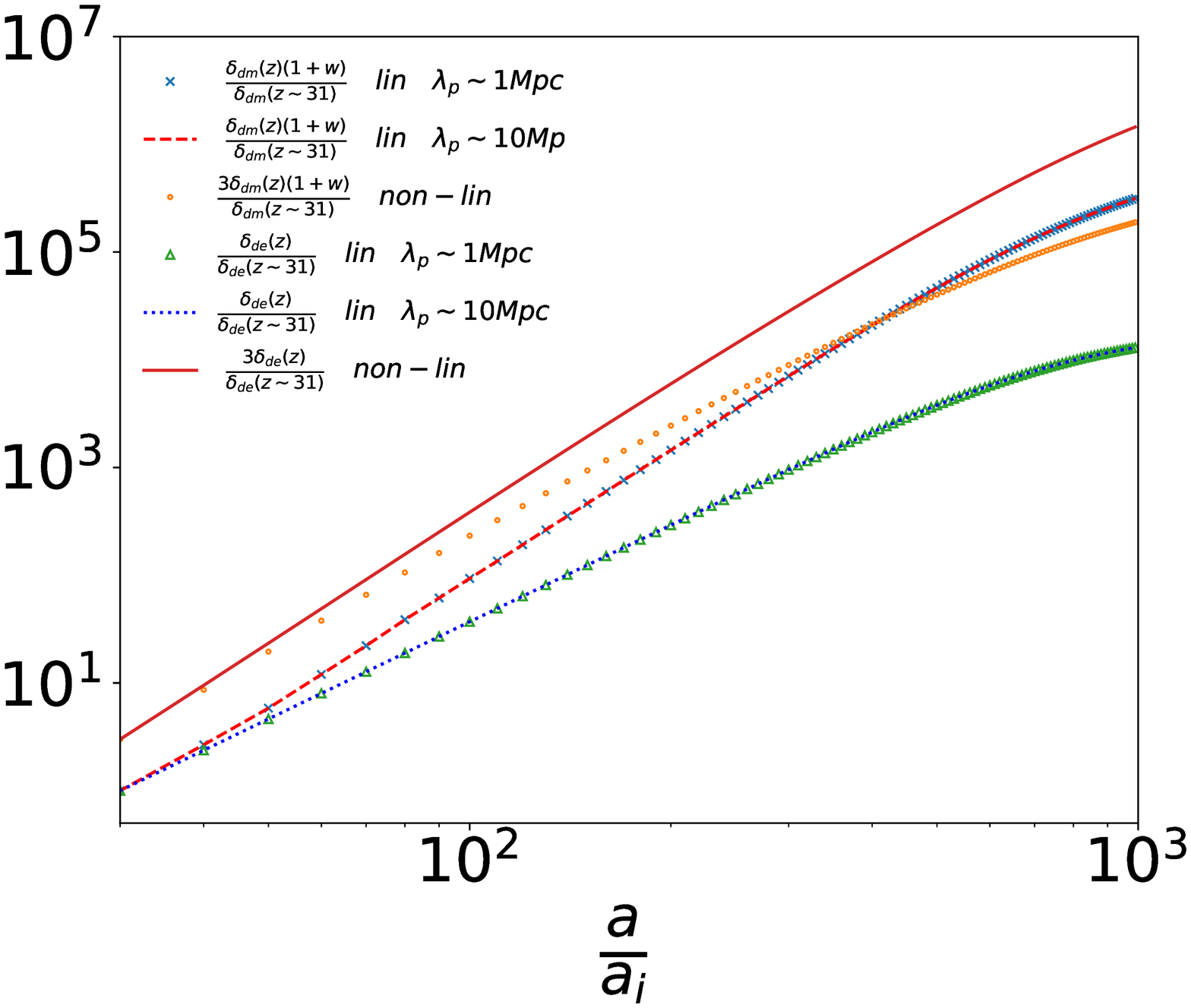}
\includegraphics[width=.45\textwidth]{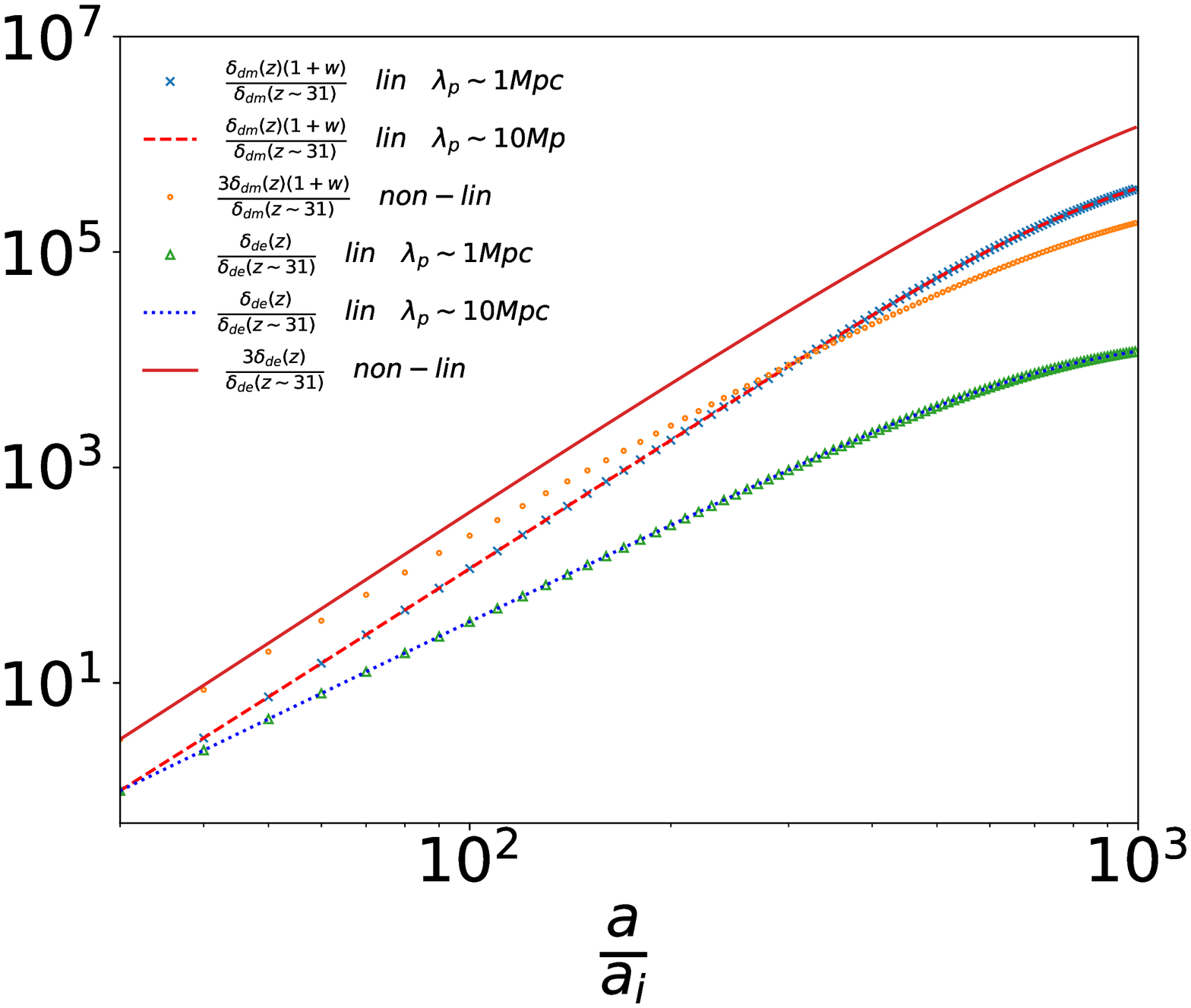}
\includegraphics[width=.45\textwidth]{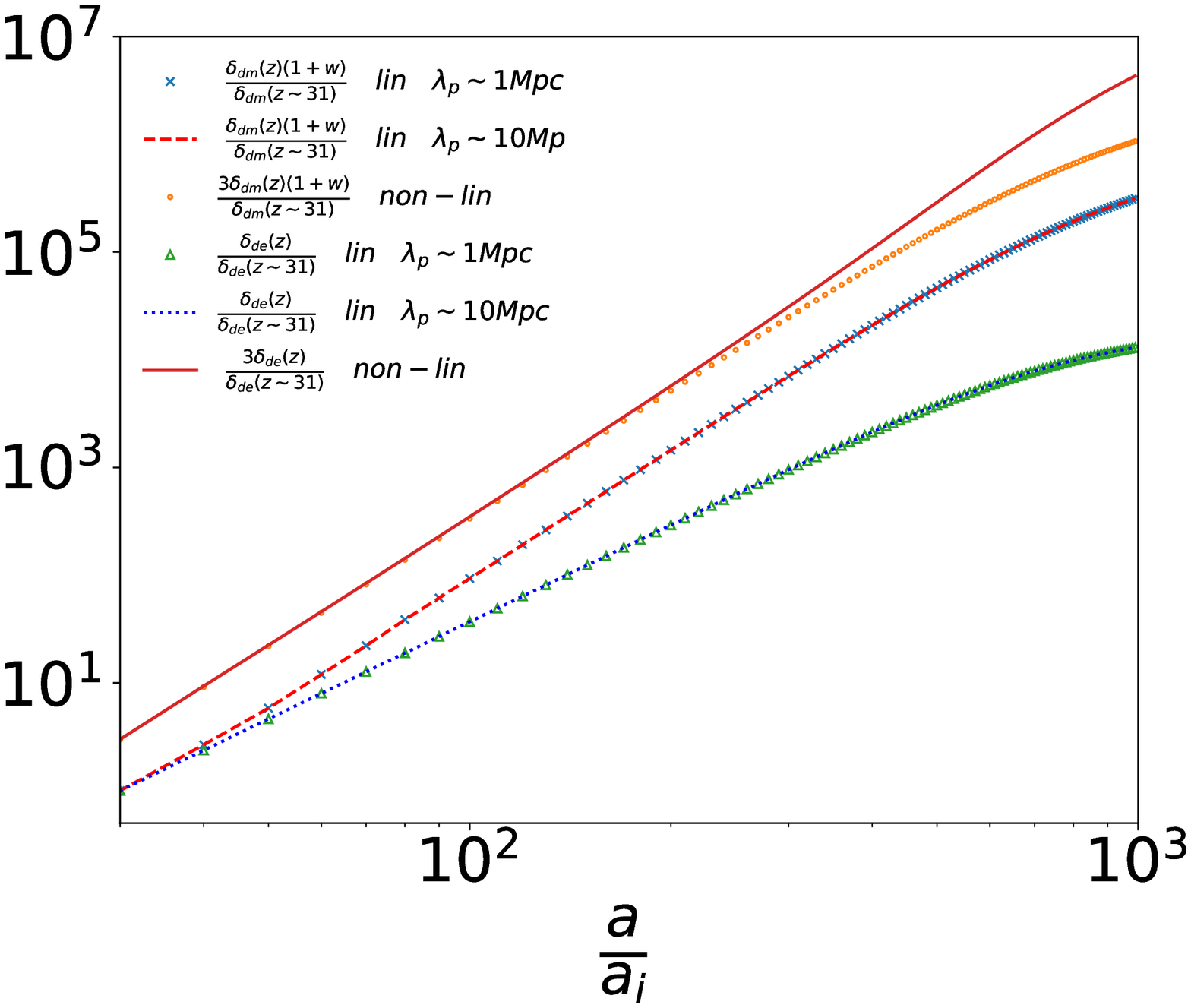}
\includegraphics[width=.45\textwidth]{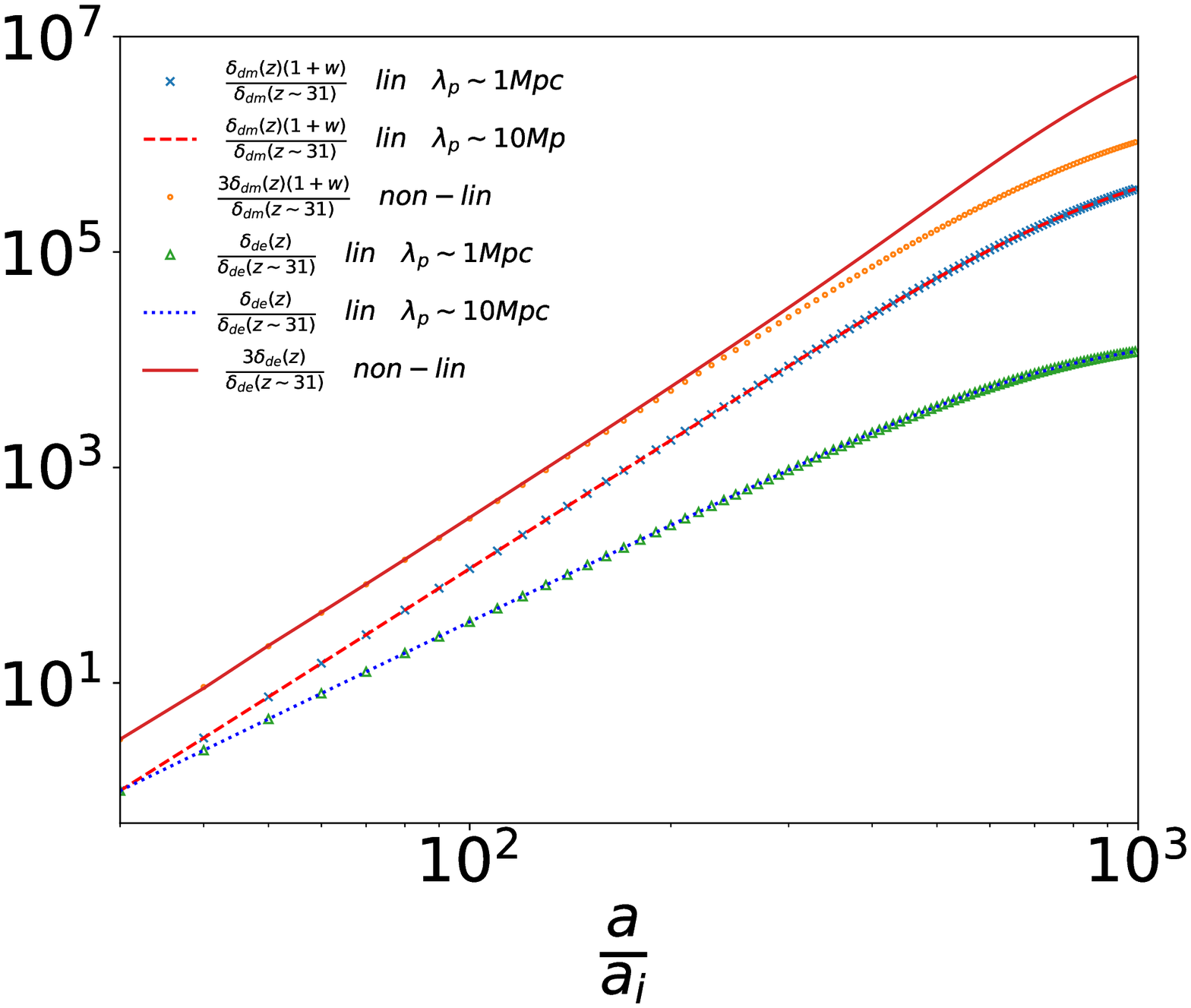}
\caption{\label{fig:14}  A comparison of the evolution of dark energy
  perturbations in our simulations UD1 (top panel) and OD1 (lower
  panel).
  Left panels are for $V\propto \psi^2 $ while right one are for
  $V\propto \exp(-\psi) $.
  At very large scales the linear theory prediction for the
  magnitude of dark energy perturbations scales as $(1+w)\delta_{dm}$.
  We have plotted this combination for linearly evolved $\delta_{dm}$
  for two scales: $1$~Mpc (cross) and $10$~Mpc (dashed line).  Linear
  evolution of dark energy density contrast for the two scales is also
  shown here as triangles ($1$~Mpc) and dotted line ($10$~Mpc).  We
  find that the linear evolution for dark energy perturbations is
  slower at small scales as compared to the expected variation at
  large scales.
  All points pertaining to linear evolution are normalised to unity at
  the left corner.
  Also shown in each panel is the combination $\delta_{dm}(1+w)$ at a
  scale of $6$~Mpc in OD1 and $20$~Mpc in UD1 simulations as small
  circles. 
  We compare this with the evolution of dark energy density contrast
  at the same scale in simulations (solid curve).
  These two sets of points are anchored at $3$ on the left corner:
  this has been done to facilitate comparison and avoid crowding.
  We find that the rate of growth of dark energy perturbations in
  simulations is higher than the linear rate, and also the combination
  $\delta_{dm}(1+w)$.  Thus dark energy perturbations grow at a faster
  rate in presence of non-linear dark matter perturbations.}
\end{figure}

\subsubsection{Comparison with Linear Perturbation Theory}

We have seen that the density contrast in dark energy remains much
smaller than unity in all cases considered here.
This makes it possible to consider density fluctuations in dark
energy at a perturbative level.
We compare the rate of growth of dark energy perturbations in our
simulations with the rate of growth expected in linear perturbation
theory.
Such a comparison is useful as it allows us to assess the significance
of non-linear dark matter perturbations that our model takes into
account.

Before carrying out the comparison, we note that the growth of
dark energy perturbations has been studied and it has been found that
the growth of perturbations is stunted at small scales.
It has been shown that at very large scales $\delta_{DE} \propto (1+w)
\delta_{DM}$, which is the expected relation for adiabatic
perturbations.  
For thawing models, $w\simeq -1$ at early times and increases slowly
over time.
Thus the rate of growth of dark energy perturbations in such models
can be much larger than the rate of growth of perturbations of dark
matter perturbations.
However, same studies indicate that the rate of growth of dark energy
perturbations at small scales is slower than the rate at large
scales.
Specifically, it has been shown that at scales much smaller than the
Hubble radius, the linear growth rate is independent of scale.

In figure ~\ref{fig:14}, we show the growth in density contrast for a
particular co-moving radius for non-linear spherical case and the
corresponding Fourier space amplitude ($\delta_k$) for two length
scales $1$~Mpc and $10$~Mpc.
We show results from two simulations: OD1 (upper panel) and UD1 (lower
panel).
The curves are normalised at the left corner to avoid crowding and
facilitate comparison. 
This also subsumes an offset required due to different initial
conditions (growing mode vs. comoving initial conditions for the two
calculations) used in the two different calculations. 
We find that the rate of growth in the two calculations differs.
In particular, at late times, the growth rate of density perturbations
in dark energy in the simulation increases and the final amplification
factor is more than a factor of ten higher than expected in the linear
perturbation theory.
In particular, we find that the dark energy density contrast grows
faster than the combination $\delta_{dm}(1+w)$ in the simulation
whereas the expectation from linear calculation is for a slower growth
rate. 
Thus the non-linear evolution of density fluctuations in dark matter
leads to a more rapid growth of perturbations in dark energy.

\begin{figure}[tbp]
\centering 
\includegraphics[width=.45\textwidth]{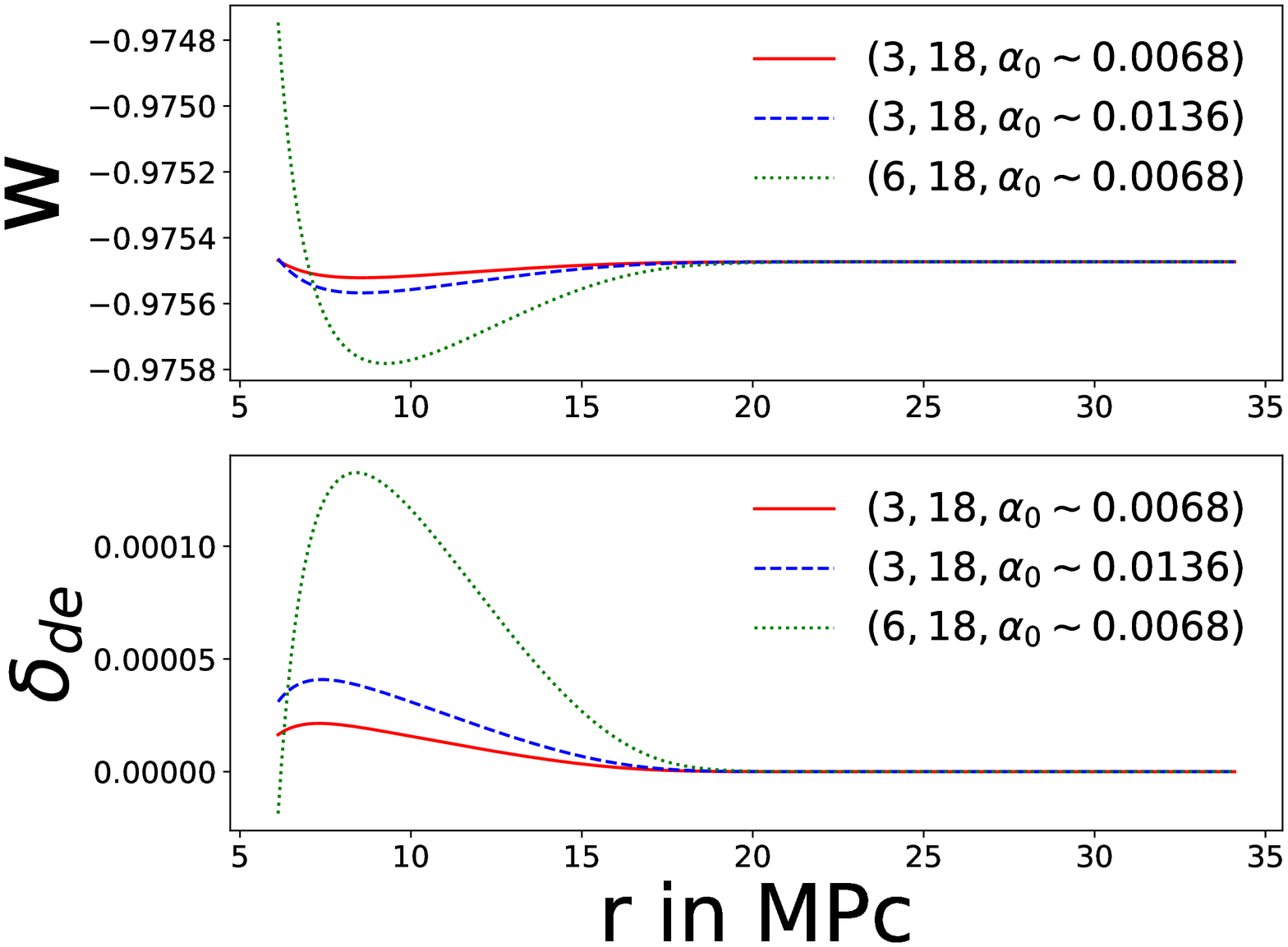}
\includegraphics[width=.45\textwidth]{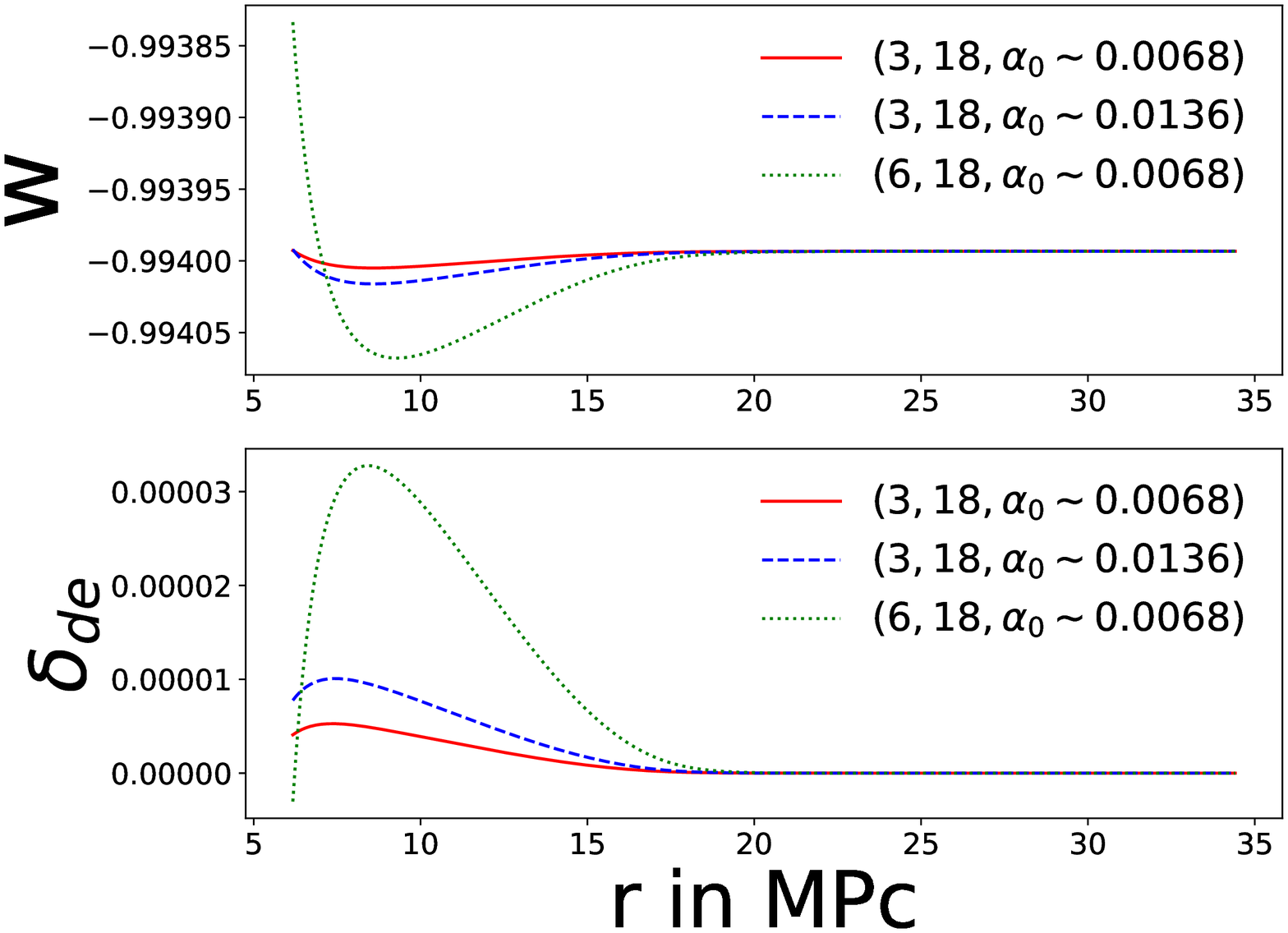}
\caption{\label{fig:b78}  In this figure we explore the dependence of
  the variation of $w$ on the scale of perturbations in dark matter
  and also on the amplitude of initial density contrast for dark
  matter.  We plot the variation of $w$ with scale for three models
  (OD1, OD2 and OD3).
  Two of the models are for the same initial density contrast in
  matter but for different scales of perturbation.  The third
  model has the same scale of perturbation as our fiducial model, but
  has a significantly higher amplitude of the initial matter
  perturbation.  We find that
  the variation of $w$ is strongest in the model with the larger scale
  but same amplitude as the fiducial model.  The variation with the change in
  amplitude of perturbation is much smaller.   The left
  panel is for $V\propto \psi^2 $ while the right panel is for
  $V\propto \exp(-\psi) $. }
\end{figure}

\begin{figure}[tbp]
\centering 
\includegraphics[width=.45\textwidth]{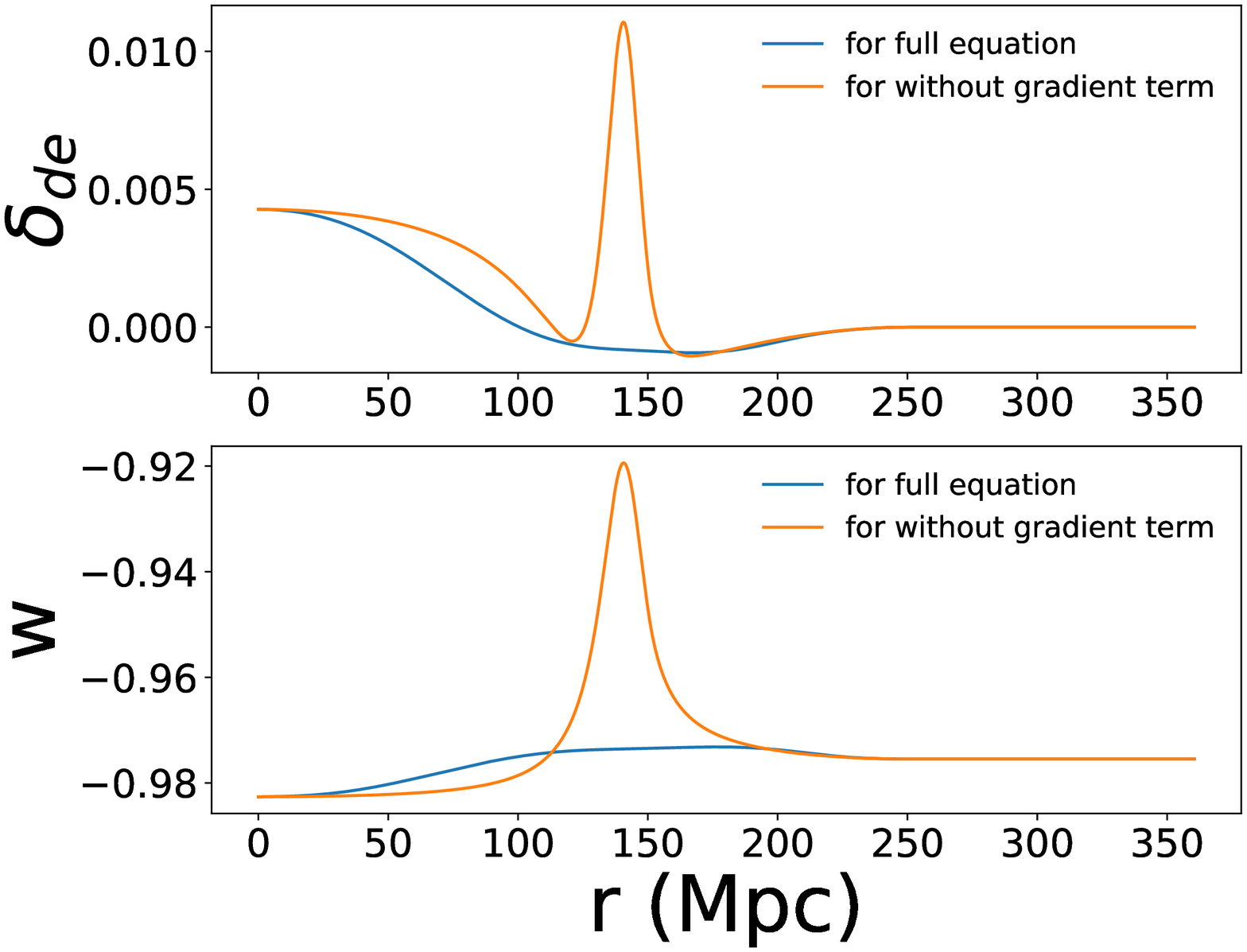}
\includegraphics[width=.45\textwidth]{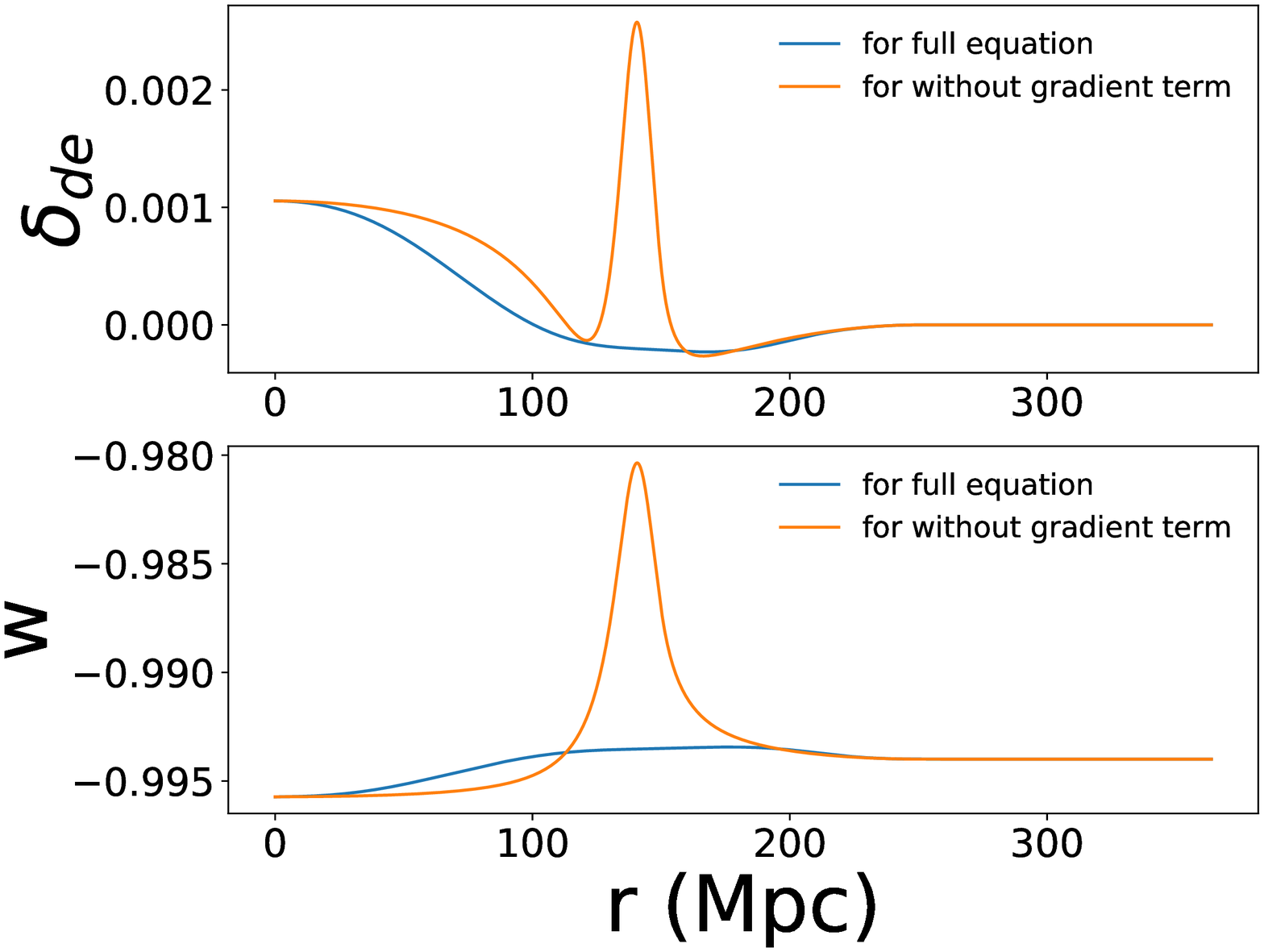}
\caption{\label{fig:bb78}  In this figure we explore the leading cause
  of variation of equation of state parameter $w$ for simulation UD1.
  We show the 
  variation computed by retaining only the local Hubble expansion
  terms in the equation of motion and compare it with the full
  simulation.  In the former case, we ignore the gradient term.  We
  find that the variation of $w$ is fairly strong and has some
  localised features when the gradient terms are ignored.  The
  localised features are not present in the full simulation
  indicating that the gradients of the scalar field are suppressed in
  the evolution, and the local Hubble expansion is not the only
  determining factor. 
}
\end{figure}

\begin{figure}[tbp]
\centering 
\includegraphics[width=.45\textwidth]{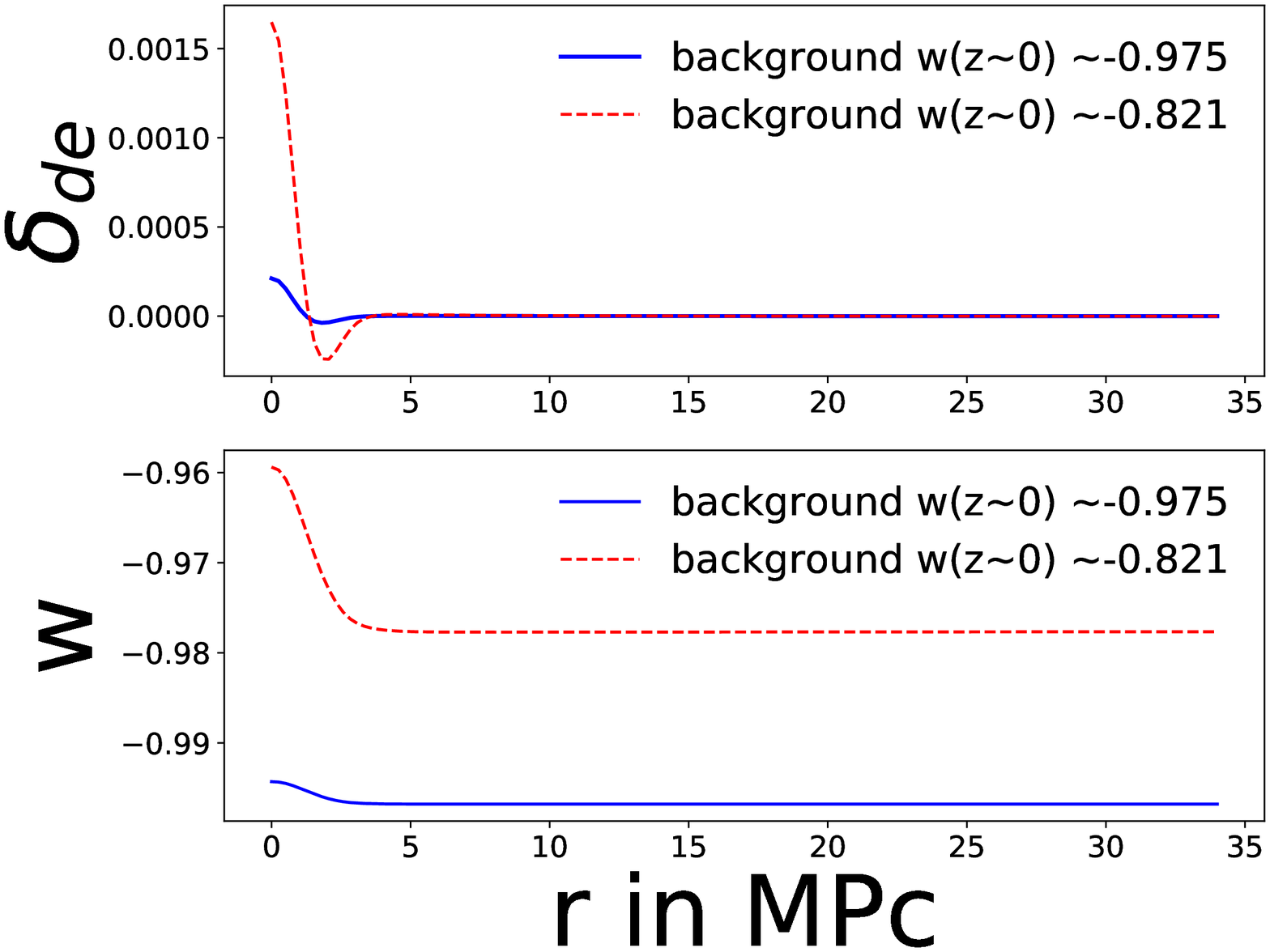}
\includegraphics[width=.45\textwidth]{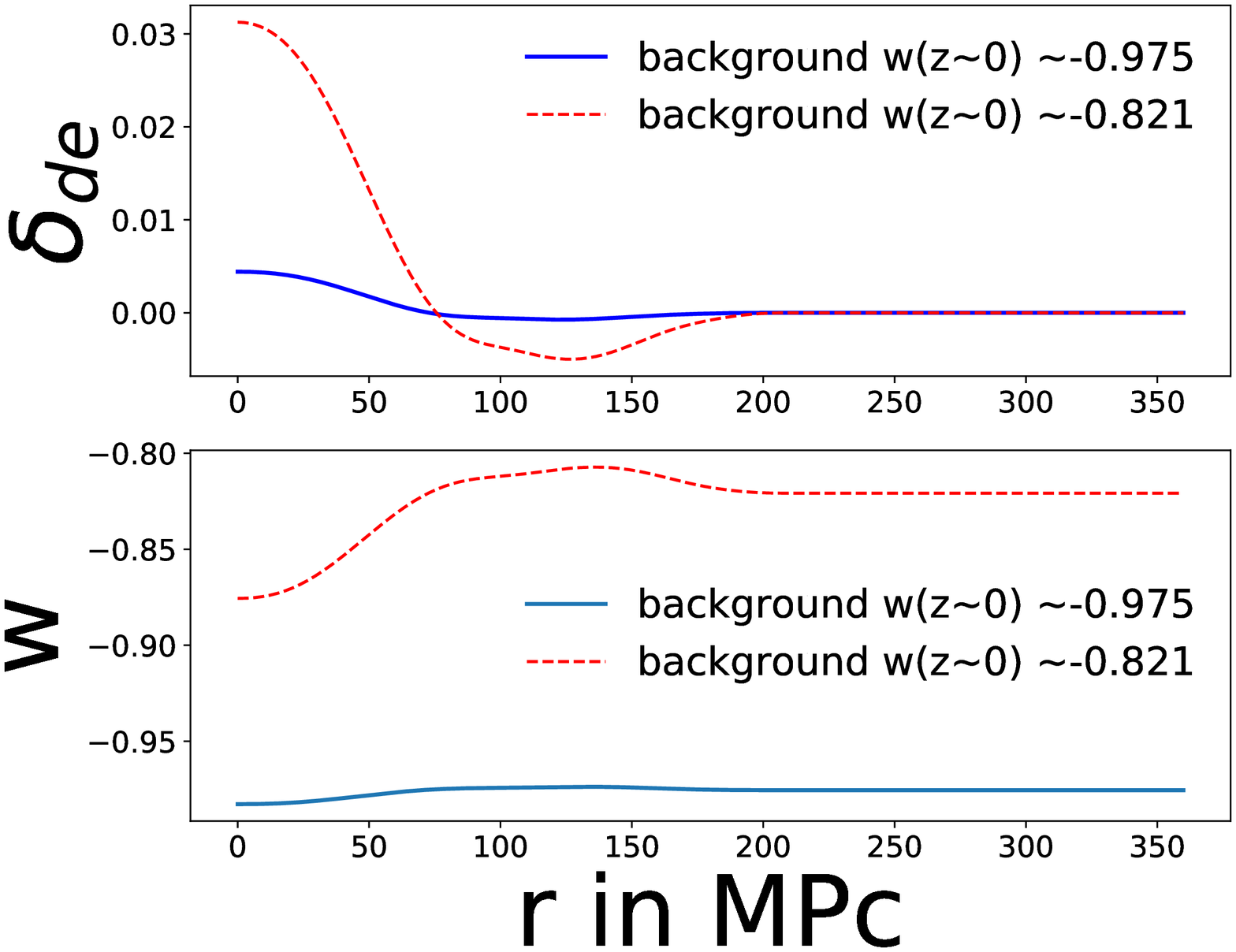}
\caption{\label{fig:15} In this figure we study the impact of the
  equation of state parameter $w$ for the background on the growth of
  dark energy perturbations and the radial variation of the equation
  of state parameter. Here we show perturbation growth in two
  different background models for $V\propto \psi^2 $. Curves are
  labeled by present day values of $w$ for the 
  background model.  We see that the perturbations have a larger
  amplitude and $w$ has a larger variation for a larger $1+w_0$.  The
  left panel here is for an over-density (OD1) and the right panel is
  for a void (UD5).  We see that the effect is strongly pronounced for
  under density partly because larger scales are involved.  The curves
  for over density are plotted for $z=1.5$, before virialisation of the
  innermost shells.  Curves for UD5 are plotted at $z=0$.}  
\end{figure}

\subsubsection{Exploring dark energy perturbations}

The variation in $w$ around a dark matter over density is caused
mainly by the slower expansion rate that leads to a more rapid rolling
down of the scalar field.
In case of under dense regions, the faster expansion slows down the
rolling of the field further.  
We test this conjecture by running a simulation with only the local
Hubble flow terms retained.
The field equation in this case reduces to:
\begin{equation}
\ddot{\psi} = -c^2\frac{\partial V}{\partial \psi}  -
  \left(\dot{B}+\frac{2\dot{R}}{R}\right) \dot{\psi} 
\end{equation}
where we have dropped the terms related to $\psi'$ and $\psi''$. 
We find that evolving the system with this equation gives rise to
sharp features that are not seen in the full simulation as shown in
Figure~\ref{fig:bb78}.
We surmise that in addition to the variation in expansion rate, there
is also a suppression of gradient of the scalar field. 

The variation of $w$ around matter perturbations is interesting and we
investigate it further.
This is important in order to ascertain the possibility of
constraining models using observations. 
Specifically, we explore the magnitude of variation as a function of
the amplitude of perturbation, i.e., $\delta_i$, and also as a
function of the scale of perturbation.

We find that the effect of the scale of perturbation is much more
important than the effect of the amplitude of initial perturbation in
matter. In Figure \ref{fig:b78}, we see that for two perturbations
with the same amplitude, variation of the scale of perturbation has
more pronounced effect than variation of amplitude of perturbation
for same scale of perturbation.

We note that we have explored the parameter space for models in the
vicinity of the cosmological constant by requiring $w \sim -1$ through
the evolution.
In models that deviate strongly from the cosmological constant, the
perturbations in dark energy become more significant.
Example in figure \ref{fig:15} illustrates this where we compare
perturbations in models with different present values of the equation
of state parameter $w(z=0)$.
The amplitude of density contrast in dark energy as well as the radial
variation in $w$ is much stronger in the model with the larger $1+w$. 
We also see that the spatial variation of $w$ for the void is very
significant when the present day value for the background deviates
strongly from $-1$.

\section{Discussion}
\label{sec:summary}

We have presented results of our analysis of spherical collapse of
dark matter and dark energy for a canonical scalar field model.
In this section we review these results, discuss the implications and
also future directions for our study.

We find that the evolution of dark matter density contrast as well as
turn around radius and virialisation are unaffected by dark energy
perturbations.
This is demonstrated by comparing the evolution in our model with an
equivalent model with the same background evolution and no
perturbations in dark energy.
This result provides justification for ignoring the role of dark
energy perturbations while studying collapse of dark matter
perturbations.
This also implies that there is no significant effect of dark energy
perturbations on structure formation.
Such effects have been studied earlier in effective models, e.g.,
\cite{2017JCAP...11..048B}.  

We have shown that the evolution of dark energy perturbations outside
the virial radius is insensitive to the scheme used to evolve dark
energy perturbations within the virial radius.
We have used the approximation of treating the scalar field as a test
field inside the virial radius, patching up with the self consistent
evolution outside the virial radius.
In all plots we have either restricted ourselves to epochs prior to
virialisation, or we have plotted functions at scales larger than the
virial radius. 

We find that the dark energy perturbations remain small, i.e.,
$|\delta_{DE}| \ll 1$ at all scales and times.
This is not to say that there is no effect of non-linear evolution of
dark matter perturbations.
We show, by comparing with the expected linear growth rate for dark
energy perturbations, that the rate of growth of dark energy
perturbations is strongly enhanced in the vicinity of non-linear dark
matter perturbations.
This is encouraging and we plan to study collapse in other dark energy
models to explore if dark energy perturbations grow to a significant
amplitude in some cases.

This finding encourages us to explore approximations between linear
perturbation theory and full non-linear collapse, where we may
consider the dark energy to have small perturbations but dark matter
may be allowed to have large over densities.
It may be possible to relax the restriction of spherical collapse in
a suitable approximation scheme. 

The most remarkable finding of our work is that the equation of state
parameter $w$ becomes a function of space.
This has been reported for fluid based models where the equation of
state parameter and the effect speed of sound for dark energy
perturbations are not the same \cite{2008PhRvD..77f7301A}.

The evolution of $w$ in the models being studied here shows a steady
increase from the initial value that is close to $-1$ for the
background.  
In the vicinity of over dense regions, this value increases at a
faster rate as the local Hubble expansion is slowed down and halted.
In voids, the local Hubble expansion is faster than the background and
the change in $w$ away from $w=-1$ is slowed down.
As a result, $w$ takes on larger values around collapsed halos and it
takes on smaller values in voids.
Thus $w$ becomes a weak function of over density and we get an
interesting coupling between the non-relativistic matter and dark
energy sectors even though we are working with a model with minimal
coupling.  

We find that the effects of dark energy clustering and spatial
variation of $w$ are strongest for large scale perturbations.  
Thus the largest over-densities and voids may be appropriate places to
look for observational evidence. 

We have considered two thawing models here but we expect that the
variation of $w$ will have an opposite trend for freezing models,
i.e., it will take on values closer to $-1$ around collapsed halos and
values away from this in voids.
This expectation follows from the evolution of the field towards the
asymptotic value of $-1$, which is slowed down or hastened by the
variations in local Hubble flow. 

We may model the relation of equation of state parameter as:
\begin{equation}
  w\left(\mathbf{r},t\right) = \bar{w} (t) + \epsilon f
\left(\delta_{DM}, \ldots \right)
\end{equation}
where $\bar{w}$ is the value for the background model, $\epsilon$ is a
small number, and $f$ is a suitable function of density contrast and
possibly other quantities such as the velocity field.
Such models can be used to explore the impact of spatial variation in
$w$ on weak lensing and other physical quantities of interest.
It may also be possible to test for such variations by stacking over
many objects/voids.
We are studying potential avenues for testing the variation of $w$ in
space. 

We are studying spherical collapse for other models of dark energy.
The motivation for these studies is to see whether we can
differentiate between such models on the basis of perturbations even
if the background evolution is identical.
We are aware of the fact that for any given form of the Lagrangian, we
can tune the potential to produce a suitable expansion history in the
form of an $a(t)$, where $a$ is the scale factor.
However, if we fix $a(t)$ then we have precisely one model for each
form of Lagrangian.
A comparison of perturbations in models with the same expansion
history will allow us to explore the information that we may extract
from observational probes of perturbations in dark energy. 

\section*{Acknowledgements}

The authors acknowledge the use of the HPC facility at IISER Mohali
for this work.
This research has made use of NASA's Astrophysics Data
System Bibliographic Services.
Authors thank Aseem Paranjape, H K Jassal and Avinash Singh for
discussions.



\appendix

\section{$T^{\mu}_ {\nu}$ for scalar field}
\label{sec:tforfield}

In order to get Einstein's equation in the familiar form, we define
the stress-energy tensor as follows:
\begin{equation}
T_{\mu \nu} = -2c\left[  \frac{\partial L_\psi}{\partial g^{\mu \nu}} 
  - \frac{1}{2} L_\psi g_{\mu \nu} \right]       \label{eq:Q5} 
\end{equation} 
Owing to spherical symmetry we get the following non-vanishing
components.
\begin{equation}
T_{\mu \nu} = c\left[ \partial_{\mu} \psi \partial_{\nu} \psi \,-\,L_\psi g_{\mu \nu}  \right]   \label{eq:Q6}
\end{equation}
\begin{equation}
T_0^0 =  c\left[ \frac{\dot{\psi}^2}{2c^2} + \frac{e^{-2B}\psi'^2}{2} + V  \right] \label{eq:Q7}
\end{equation}
\begin{equation}
T_1^1 =  -c\left[ \frac{\dot{\psi}^2}{2c^2} + \frac{e^{-2B}\psi'^2}{2} - V  \right] \label{eq:Q8}
\end{equation}
\begin{equation}
T_2^2 = T_3^3 =  -c\left[ \frac{\dot{\psi}^2}{2c^2} - \frac{e^{-2B}\psi'^2}{2} - V  \right]
\label{eq:Q9}
\end{equation}
\begin{equation}
T_0^1 = -ce^{-2B}\dot{\psi}\psi'  \label{eq:Q10}
\end{equation}
\begin{equation}
T_1^0 = \frac{\dot{\psi}\psi'}{c}  \label{eq:Q11}
\end{equation}

Vanishing of four divergence of stress energy tensor gives us the
equation of motion for the scalar field:
\begin{equation}
{T_0^\mu,}_\mu = c\dot{\psi}\left[  e^{-2B}\left(B'\psi' - \psi'' - 2\frac{R'}{R}\psi' \right) + \frac{\dot{B}\dot{\psi}}{c^2} + 
\frac{2\dot{\psi}\dot{R}}{Rc^2} + \frac{\ddot{\psi}}{c^2} + {V,}_\psi\right] = 0   \label{eq:Q13}
\end{equation}

\section{Einstein Equations} 
\label{sec:ee}

Variation of Einstein-Hilbert action gives us:
\begin{equation*}
\delta I_{Ein-Hilb} = \frac{c^3}{16 \pi G}\int{(dr d\theta d\phi dt) \sqrt{-g}\left[R_{\mu \nu} - \frac{1}{2} g_{\mu \nu} R_E\right] 
\delta g^{\mu \nu}}   \label{eq:Q15}
\end{equation*}
where Ricci scalar is represented as $R_E$ to distinguish it from
metric coefficient $R$. 
Combining this variation with the stress- energy tensor for $\psi$ in
previous sub-subsection, we get Einstein's equations 
\begin{equation*}
G^\mu_\nu = R^\mu_\nu - \frac{1}{2} \delta^\mu_\nu R_E = \frac{8\pi G}{c^4} T^\mu_\nu  \label{eq:Q16}
\end{equation*} 
$\left(^1_1\right)$ component
\begin{equation}
\left[ \frac{1}{R^2} - e^{-2B}\frac{R'^2}{R^2} + \frac{\dot{R}^2}{c^2 R^2} + \frac{2\ddot{R}}{c^2 R}\right]
= -\frac{8\pi G}{c^3}\left[ \frac{\dot{\psi}^2}{2c^2} + \frac{e^{-2B}\psi'^2}{2} - V  \right] 
\label{eq:Q17}
\end{equation}
$\left(^2_2\right)$ and $\left(^3_3\right)$ component
\begin{equation}
e^{-2B} \left[ \frac{R'B'}{R} - \frac{R''}{R} \right] + \frac{1}{c^2} 
\left[ \frac{\dot{R}\dot{B}}{R} + \frac{\ddot{R}}{R} + \dot{B}^2 + \ddot{B} \right]
= -\frac{8\pi G}{c^3} \left[ \frac{\dot{\psi}^2}{2c^2} - \frac{e^{-2B}\psi'^2}{2} - V  \right] 
\label{eq:Q18}
\end{equation}
$\left(^0_0\right)$  component
\begin{equation}
\begin{split}
-e^{-2B} \left[ \left( \frac{R'}{R}\right)^2  - \frac{2R'B'}{R} + \frac{2R''}{R} \right]
+ \frac{1}{R^2} + \frac{\dot{R}^2}{c^2 R^2} + \frac{2\dot{R}\dot{B}}{c^2 R} 
= \\
 \frac{8\pi G \rho}{c^2} + \frac{8\pi G}{c^3}  
\left[ \frac{\dot{\psi}^2}{2c^2} + \frac{e^{-2B}\psi'^2}{2} + V  \right] 
\end{split}   \label{eq:Q19}
\end{equation}
$\left(^1_0\right)$ and $\left(^0_1\right)$ components yield same equation
\begin{equation}
R'\dot{B} - \dot{R}' = \frac{4\pi G}{c^3}\dot{\psi}\psi' R      \label{eq:Q20}
\end{equation}
Combining equations for $\left(^0_0\right)$,$\left(^1_1\right)$ and
$\left(^2_2\right)$ components, we obtain: 
\begin{equation}
\ddot{B} = \frac{8\pi G}{c}\left[ e^{-2B}\psi'^2 + V + \frac{\rho c}{2} \right]   + 2e^{-2B}c^2\left[ \frac{R''}{R} - \frac{R'B'}{R} \right]
- \frac{2\dot{B}\dot{R}}{R} - \dot{B}^2     \label{eq:Q21}
\end{equation}
or equivalently we can obtain
\begin{equation}
\ddot{B} = -c^2e^{-2B}\frac{R'^2}{R^2} + \frac{c^2}{R^2} + \frac{\dot{R}^2}{R^2} - \dot{B}^2 - 4\pi G\rho 
- \frac{8\pi G}{c} \left[ \frac{\dot{\psi}^2}{2c^2}  - e^{-2B}\frac{\psi'^2}{2} \right] 
\label{eq:Q22}
\end{equation}
and from $\left(^1_1\right)$, we already have
eqn.\eqref{eq:Q17}. Rewriting it again 
\begin{equation}
\frac{\ddot{R}}{R} = -\frac{4\pi G}{c}\left[ \frac{\dot{\psi}^2}{2c^2} + \frac{e^{-2B}\psi'^2}{2} - V  \right]
-\frac{1}{2}\frac{\dot{R}^2}{R^2} + \frac{c^2}{2} \left[  e^{-2B}\frac{R'^2}{R^2} - \frac{1}{R^2} \right]   \label{eq:Q23}
\end{equation}

\section{Numerical Methods}
\label{Numerix}

Three second order partial differential equations,
eq~\eqref{eq:Q26},\eqref{eq:Q27} and \eqref{eq:Q14}, can be written as
6 first order partial differential equations and we have two first
order partial differential equations for $\rho$  and $R'$ giving us
total of 8 first order partial differential equations. 
\begin{eqnarray}
\dot{x}_i(r) = f_i\left[
  x_1(r),x_2(r),..,x_8(r),x'_1(r),x''_1(r),x'_3(r)\right]\\ 
\left\lbrace x_1,x_2,x_3,x_4,x_5,x_6,x_7,x_8\right\rbrace  =
  \left\lbrace \psi,R,B,\dot{\psi},\dot{R},\dot{B},\rho,R'
  \right\rbrace 
\end{eqnarray}
But solving these equations using time 't' as parameter turns out to
be time consuming, so we switch to background scale factor 'a(t)' as
independent parameter. Switching from 't' to 'a' requires
simultaneously solving two more equations for $\dot{a}$ and
$\ddot{a}$: 
   
Having structured equations in above form, we used a RK4 algorithm to
solve the equations in following flow: 
\begin{itemize}
\item
Initialise all variables
\item
Loop over "a" begins
\subitem
Calculate spatial derivatives
\subitem
RK4 first predictor step to calculate $x_ik1$'s
\subitem
Calculate spatial derivatives
\subitem
RK4 second predictor step to calculate $x_ik2$'s
\subitem
Calculate spatial derivatives
\subitem
RK4 third predictor step to calculate $x_ik3$'s
\subitem
Calculate spatial derivatives
\subitem
RK4 fourth predictor step to calculate $x_ik4$'s and correction.
\item
Loop over "a" ends
\end{itemize}

We have tested the algorithm for numerical convergence by varying
$\Delta t$ and $\Delta r$.
Further, the epoch of virialisation scales
correctly with initial density contrasts.
We have also solved the equations in the case of $\Lambda$CDM and
compared with the solutions obtained using the first integral. 
These tests have been used to validate the code.

\bibliographystyle{plain}


\begin{thebibliography}{10}

\bibitem{1990Natur.348..705E}
G.~{Efstathiou}, W.~J. {Sutherland}, and S.~J. {Maddox}.
\newblock {The cosmological constant and cold dark matter}.
\newblock {\em \nat}, 348:705--707, December 1990.

\bibitem{1993Natur.366..429W}
S.~D.~M. {White}, J.~F. {Navarro}, A.~E. {Evrard}, and C.~S. {Frenk}.
\newblock {The baryon content of galaxy clusters: a challenge to cosmological
  orthodoxy}.
\newblock {\em \nat}, 366:429--433, December 1993.

\bibitem{1995Natur.377..600O}
J.~P. {Ostriker} and P.~J. {Steinhardt}.
\newblock {The observational case for a low-density Universe with a non-zero
  cosmological constant}.
\newblock {\em \nat}, 377:600--602, October 1995.

\bibitem{1996ComAp..18..275B}
J.~S. {Bagla}, T.~{Padmanabhan}, and J.~V. {Narlikar}.
\newblock {Crisis in Cosmology: Observational Constraints on {$\Omega$} and H
  $_{0}$}.
\newblock {\em Comments on Astrophysics}, 18:275, 1996.

\bibitem{1998AJ....116.1009R}
A.~G. {Riess}, A.~V. {Filippenko}, P.~{Challis}, A.~{Clocchiatti},
  A.~{Diercks}, P.~M. {Garnavich}, R.~L. {Gilliland}, C.~J. {Hogan}, S.~{Jha},
  R.~P. {Kirshner}, B.~{Leibundgut}, M.~M. {Phillips}, D.~{Reiss}, B.~P.
  {Schmidt}, R.~A. {Schommer}, R.~C. {Smith}, J.~{Spyromilio}, C.~{Stubbs},
  N.~B. {Suntzeff}, and J.~{Tonry}.
\newblock {Observational Evidence from Supernovae for an Accelerating Universe
  and a Cosmological Constant}.
\newblock {\em \aj}, 116:1009--1038, September 1998.

\bibitem{1999ApJ...517..565P}
S.~{Perlmutter}, G.~{Aldering}, G.~{Goldhaber}, R.~A. {Knop}, P.~{Nugent},
  P.~G. {Castro}, S.~{Deustua}, S.~{Fabbro}, A.~{Goobar}, D.~E. {Groom}, I.~M.
  {Hook}, A.~G. {Kim}, M.~Y. {Kim}, J.~C. {Lee}, N.~J. {Nunes}, R.~{Pain},
  C.~R. {Pennypacker}, R.~{Quimby}, C.~{Lidman}, R.~S. {Ellis}, M.~{Irwin},
  R.~G. {McMahon}, P.~{Ruiz-Lapuente}, N.~{Walton}, B.~{Schaefer}, B.~J.
  {Boyle}, A.~V. {Filippenko}, T.~{Matheson}, A.~S. {Fruchter}, N.~{Panagia},
  H.~J.~M. {Newberg}, W.~J. {Couch}, and T.~S.~C. {Project}.
\newblock {Measurements of {$\Omega$} and {$\Lambda$} from 42 High-Redshift
  Supernovae}.
\newblock {\em \apj}, 517:565--586, June 1999.

\bibitem{1998ApJ...507...46S}
B.~P. {Schmidt}, N.~B. {Suntzeff}, M.~M. {Phillips}, R.~A. {Schommer},
  A.~{Clocchiatti}, R.~P. {Kirshner}, P.~{Garnavich}, P.~{Challis},
  B.~{Leibundgut}, J.~{Spyromilio}, A.~G. {Riess}, A.~V. {Filippenko},
  M.~{Hamuy}, R.~C. {Smith}, C.~{Hogan}, C.~{Stubbs}, A.~{Diercks}, D.~{Reiss},
  R.~{Gilliland}, J.~{Tonry}, J.~{Maza}, A.~{Dressler}, J.~{Walsh}, and
  R.~{Ciardullo}.
\newblock {The High-Z Supernova Search: Measuring Cosmic Deceleration and
  Global Curvature of the Universe Using Type IA Supernovae}.
\newblock {\em \apj}, 507:46--63, November 1998.

\bibitem{2000ApJ...536L..63M}
A.~{Melchiorri}, P.~A.~R. {Ade}, P.~{de Bernardis}, J.~J. {Bock}, J.~{Borrill},
  A.~{Boscaleri}, B.~P. {Crill}, G.~{De Troia}, P.~{Farese}, P.~G. {Ferreira},
  K.~{Ganga}, G.~{de Gasperis}, M.~{Giacometti}, V.~V. {Hristov}, A.~H.
  {Jaffe}, A.~E. {Lange}, S.~{Masi}, P.~D. {Mauskopf}, L.~{Miglio}, C.~B.
  {Netterfield}, E.~{Pascale}, F.~{Piacentini}, G.~{Romeo}, J.~E. {Ruhl}, and
  N.~{Vittorio}.
\newblock {A Measurement of {$\Omega$} from the North American Test Flight of
  Boomerang}.
\newblock {\em \apjl}, 536:L63--L66, June 2000.

\bibitem{2003ApJS..148..175S}
D.~N. {Spergel}, L.~{Verde}, H.~V. {Peiris}, E.~{Komatsu}, M.~R. {Nolta}, C.~L.
  {Bennett}, M.~{Halpern}, G.~{Hinshaw}, N.~{Jarosik}, A.~{Kogut}, M.~{Limon},
  S.~S. {Meyer}, L.~{Page}, G.~S. {Tucker}, J.~L. {Weiland}, E.~{Wollack}, and
  E.~L. {Wright}.
\newblock {First-Year Wilkinson Microwave Anisotropy Probe (WMAP) Observations:
  Determination of Cosmological Parameters}.
\newblock {\em \apjs}, 148:175--194, September 2003.

\bibitem{2016AA...594A..13P}
{Planck Collaboration}, P.~A.~R. {Ade}, N.~{Aghanim}, M.~{Arnaud},
  M.~{Ashdown}, J.~{Aumont}, C.~{Baccigalupi}, A.~J. {Banday}, R.~B.
  {Barreiro}, J.~G. {Bartlett}, and et~al.
\newblock {Planck 2015 results. XIII. Cosmological parameters}.
\newblock {\em \aap}, 594:A13, September 2016.

\bibitem{2010MNRAS.405.2639J}
H.~K. {Jassal}, J.~S. {Bagla}, and T.~{Padmanabhan}.
\newblock {Understanding the origin of CMB constraints on dark energy}.
\newblock {\em \mnras}, 405:2639--2650, July 2010.

\bibitem{2014MNRAS.441.3643P}
E.~{Piedipalumbo}, E.~{Della Moglie}, M.~{De Laurentis}, and P.~{Scudellaro}.
\newblock {High-redshift investigation on the dark energy equation of state}.
\newblock {\em \mnras}, 441:3643--3655, July 2014.

\bibitem{2015MNRAS.446.1321H}
S.~{Hotchkiss}, S.~{Nadathur}, S.~{Gottl{\"o}ber}, I.~T. {Iliev}, A.~{Knebe},
  W.~A. {Watson}, and G.~{Yepes}.
\newblock {The Jubilee ISW Project - II. Observed and simulated imprints of
  voids and superclusters on the cosmic microwave background}.
\newblock {\em \mnras}, 446:1321--1334, January 2015.

\bibitem{2017arXiv171000846J}
D.~O. {Jones}, D.~M. {Scolnic}, A.~G. {Riess}, A.~{Rest}, R.~P. {Kirshner},
  E.~{Berger}, R.~{Kessler}, Y.-C. {Pan}, R.~J. {Foley}, R.~{Chornock}, C.~A.
  {Ortega}, P.~J. {Challis}, W.~S. {Burgett}, K.~C. {Chambers}, P.~W. {Draper},
  H.~{Flewelling}, M.~E. {Huber}, N.~{Kaiser}, R.-P. {Kudritzki},
  N.~{Metcalfe}, J.~{Tonry}, R.~J. {Wainscoat}, C.~{Waters}, E.~E.~E. {Gall},
  R.~{Kotak}, M.~{McCrum}, S.~J. {Smartt}, and K.~W. {Smith}.
\newblock {Measuring Dark Energy Properties with Photometrically Classified
  Pan-STARRS Supernovae. II. Cosmological Parameters}.
\newblock {\em ArXiv e-prints}, October 2017.

\bibitem{c577c8168f6e4d9c8cc679edcd67e494}
Gong-Bo Zhao, Marco Raveri, Levon Pogosian, Yuting Wang, {Robert G.}
  Crittenden, {Will J.} Handley, {Will J.} Percival, Florian Beutler, Jonathan
  Brinkmann, Chia-Hsun Chuang, {Antonio J.} Cuesta, {Daniel J.} Eisenstein,
  Francisco-Shu Kitaura, Kazuya Koyama, Benjamin L'Huillier, {Robert C.}
  Nichol, {Matthew M.} Pieri, Sergio Rodriguez-Torres, {Ashley J.} Ross,
  Graziano Rossi, {Ariel G.} Sánchez, Arman Shafieloo, {Jeremy L.} Tinker,
  Rita Tojeiro, {Jose A.} Vazquez, and Hanyu Zhang.
\newblock Dynamical dark energy in light of the latest observations.
\newblock {\em Nature Astronomy}, 8 2017.
\newblock 27 pages, 3 figures and one table. A supplementary document is
  included. The BOSS DR12 BAO data used in the work can be downloaded from the
  SDSS website.

\bibitem{2017JCAP...06..012T}
A.~{Tripathi}, A.~{Sangwan}, and H.~K. {Jassal}.
\newblock {Dark energy equation of state parameter and its evolution at low
  redshift}.
\newblock {\em \jcap}, 2017:012, June 2017.

\bibitem{2017JCAP...07..040D}
  S. Dhawan, A. Goobar, E. M{\"o}rtsell, R. Amanullah and U. Feindt.
  \newblock {\em Narrowing down the possible explanations of cosmic
    acceleration with geometric probes}.
  \newblock {\em \jcap}, 2017:40, July 2017.
  

\bibitem{2018arXiv180108553V}
  S. Vagnozzi, S. Dhawan, M. Gerbino, K. Freese, A. Goobar and
  O. Mena.
  \newblock {\em Constraints on the sum of the neutrino masses in
    dynamical dark energy models with $w(z) \geq -1$ are tighter than
    those obtained in $\Lambda$CDM}.
  \newblock {\em arXiv: 1801.08553}, January 2018.

\bibitem{1989RvMP...61....1W}
S.~{Weinberg}.
\newblock {The cosmological constant problem}.
\newblock {\em Reviews of Modern Physics}, 61:1--23, January 1989.

\bibitem{2008GReGr..40..529P}
T.~{Padmanabhan}.
\newblock {Dark energy and gravity}.
\newblock {\em General Relativity and Gravitation}, 40:529--564, February 2008.

\bibitem{2010deto.book.....A}
L.~{Amendola} and S.~{Tsujikawa}.
\newblock {\em Dark Energy: Theory and Observations}.
\newblock {\em Cambridge University Press}, 2010.

\bibitem{2012ApSS.342..155B}
  K. Bamba, S. Capozziello, S. Nojiri, S.~D.~Odintsov
  \newblock {\em Dark energy cosmology: the equivalent description via
    different theoretical models and cosmography tests}. 
  \newblock {\em \apss}, 342:155-228, November 2012 

\bibitem{2013CQGra..30u4003T}
S.~{Tsujikawa}.
\newblock {Quintessence: a review}.
\newblock {\em Classical and Quantum Gravity}, 30(21):214003, November 2013.

\bibitem{2013PhR...530...87W}
D.~H. {Weinberg}, M.~J. {Mortonson}, D.~J. {Eisenstein}, C.~{Hirata}, A.~G.
  {Riess}, and E.~{Rozo}.
\newblock {Observational probes of cosmic acceleration}.
\newblock {\em \physrep}, 530:87--255, September 2013.

\bibitem{2018RPPh...81a6901H}
D.~{Huterer} and D.~L. {Shafer}.
\newblock {Dark energy two decades after: observables, probes, consistency
  tests}.
\newblock {\em Reports on Progress in Physics}, 81(1):016901, January 2018.

\bibitem{2002PhRvD..66b1301P}
T.~{Padmanabhan}.
\newblock {Accelerated expansion of the universe driven by tachyonic matter}.
\newblock {\em \prd}, 66(2):021301, June 2002.

\bibitem{2009PhRvD..79l7301J}
H.~K. {Jassal}.
\newblock {Comparison of perturbations in fluid and scalar field models of dark
  energy}.
\newblock {\em \prd}, 79(12):127301, June 2009.

\bibitem{1922ZPhy...10..377F}
  {{Friedmann}, A.}.
  \newblock {{\"U}ber die Kr{\"u}mmung des Raumes}.
  \newblock {\em Zeitschrift fur Physik}, 10:377-385, 1922.

\bibitem{1931MNRAS..91..483L}
  {{Lema{\^i}tre}, G.}.
\newblock {Expansion of the universe, A homogeneous universe of
  constant mass and increasing radius accounting for the radial
  velocity of extra-galactic nebulae}. 
\newblock {\em \mnras}, 91:483-490, March 1931.

\bibitem{1935ApJ....82..284R}
  {{Robertson}, H.~P.}.
  \newblock {Kinematics and World-Structure}.
  \newblock {\em \apj}, 82:284, Nov.1935.

\bibitem{1935QJMat...6...81W}
  {Walker A.~G.}.
  \newblock{On Riemannian spaces with spherical symmetry about a line,
    and the conditions for isotropy in general relativity}. 
  \newblock{\em The Quarterly Journal of Mathematics}, 6:81-93, 1935.

\bibitem{1972ApJ...176....1G}
J.~E. {Gunn} and J.~R. {Gott}, III.
\newblock {On the Infall of Matter Into Clusters of Galaxies and Some Effects
  on Their Evolution}.
\newblock {\em \apj}, 176:1, August 1972.

\bibitem{1984ApJ...284..439P}
  {Peebles P.~J.~E.}.
  \newblock {Tests of cosmological models constrained by inflation}.
  \newblock {\em \apj}, 284:439-444, September 1984.

\bibitem{1996MNRAS.282..263E}
  {Eke V.~R., Cole S., Frenk C.~S.}.
  \newblock {Cluster evolution as a diagnostic for Omega}.
  \newblock {\em \mnras}, 282:263-280, September 1996.

\bibitem{1991MNRAS.251..128L}
  {Lahav O., Lilje P.~B., Primack J.~R., Rees M.~J.}.
  \newblock {Dynamical effects of the cosmological constant}.
  \newblock {\em \mnras}, 251:128-136, July 1991.

\bibitem{1993MNRAS.262..717B}
J.~D. {Barrow} and P.~{Saich}.
\newblock {Growth of large-scale structure with a cosmological constant}.
\newblock {\em \mnras}, 262:717--725, June 1993.

\bibitem{1998PhRvL..80.1582C}
R.~R. {Caldwell}, R.~{Dave}, and P.~J. {Steinhardt}.
\newblock {Cosmological Imprint of an Energy Component with General Equation of
  State}.
\newblock {\em Physical Review Letters}, 80:1582--1585, February 1998.

\bibitem{2007JCAP...11..012A}
  L. R. Abramo, R. C. Batista, L. Liberato and R. Rosenfeld.
  \newblock {\em Structure formation in the presence of dark energy
    perturbations}.
  \newblock {\em \jcap}, 2007:12, November 2007.

\bibitem{2009PhRvD..79b3516A}
  L. R. Abramo, R. C. Batista, L. Liberato and R. Rosenfeld.
  \newblock {\em Physical approximations for the nonlinear evolution
    of perturbations in inhomogeneous dark energy scenarios}.
  \newblock {\em \prd}, 79:023516, January 2009.

\bibitem{2012JCAP...01..025M}
  V. Marra and M. {P{\"a}{\"a}kk{\"o}nen}.
\newblock {\em Exact spherically-symmetric inhomogeneous model with n
  perfect fluids}.
\newblock {\em \jcap}, 2012:25, January 2012.

\bibitem{2014PhRvD..89h3002R}
  Z. Roupas, M. Axenides, G. Georgiou and E.~N.~Saridakis.
  \newblock {\em Galaxy clusters and structure formation in
    quintessence versus phantom dark energy universe}.
  \newblock {\em \prd}, 89.083002, April 2014.
  
\bibitem{2010JCAP...03..027C}
P.~{Creminelli}, G.~{D'Amico}, J.~{Nore{\~n}a}, L.~{Senatore}, and
  F.~{Vernizzi}.
\newblock {Spherical collapse in quintessence models with zero speed of sound}.
\newblock {\em \jcap}, 2010:027, March 2010.

\bibitem{2010JCAP...10..014B}
  G. Ballesteros and J. Lesgourgues.
  \newblock {\em Dark energy with non-adiabatic sound speed: initial
    conditions and detectability}.
  \newblock {\em \jcap}, 2010:014, October 2010.

\bibitem{2011JCAP...11..014A}
S.~{Anselmi}, G.~{Ballesteros}, and M.~{Pietroni}.
\newblock {Non-linear dark energy clustering}.
\newblock {\em \jcap}, 2011:014, November 2011.

\bibitem{2015AASP....5...51T}
M.~{Tsizh} and B.~{Novosyadlyj}.
\newblock {Dynamics of dark energy in collapsing halo of dark matter}.
\newblock {\em Advances in Astronomy and Space Physics}, 5:51--56, September
  2015.

\bibitem{1475-7516-2016-09-031}
Christian Fidler, Thomas Tram, Cornelius Rampf, Robert Crittenden, Kazuya
  Koyama, and David Wands.
\newblock Relativistic interpretation of newtonian simulations for cosmic
  structure formation.
\newblock {\em Journal of Cosmology and Astroparticle Physics}, 2016(09):031,
  2016.

\bibitem{PhysRevD.93.043533}
J.~Rekier, A.~F\"uzfa, and I.~Cordero-Carri\'on.
\newblock Nonlinear cosmological spherical collapse of quintessence.
\newblock {\em Phys. Rev. D}, 93:043533, Feb 2016.

\bibitem{PhysRevD.95.064029}
D.~Herrera, I.~Waga, and S.~E. Jor\'as.
\newblock Calculation of the critical overdensity in the spherical-collapse
  approximation.
\newblock {\em Phys. Rev. D}, 95:064029, Mar 2017.

\bibitem{0004-637X-841-1-63}
Shuxun Tian, Shuo Cao, and Zong-Hong Zhu.
\newblock The dynamics of inhomogeneous dark energy.
\newblock {\em The Astrophysical Journal}, 841(1):63, 2017.

\bibitem{2017PhRvD..96h3506A}
I.~{Achitouv}.
\newblock {Improved model of redshift-space distortions around voids:
  Application to quintessence dark energy}.
\newblock {\em \prd}, 96(8):083506, October 2017.

\bibitem{2018PDU....19...12C}
C.-C. {Chang}, W.~{Lee}, and K.-W. {Ng}.
\newblock {Spherical collapse models with clustered dark energy}.
\newblock {\em Physics of the Dark Universe}, 19:12--20, March 2018.

\bibitem{1934PNAS...20..169T}
  Tolman, R.~C.
  \newblock{Effect of Inhomogeneity on Cosmological Models}.
  \newblock{\em Proceedings of the National Academy of Science},
  20:169--176, March 1934.

\bibitem{1947MNRAS.107..410B}
  Bondi, H.
  \newblock {Spherically symmetrical models in general relativity}.
  \newblock {\em \mnras}. 107:410--425, 1947.

\bibitem{2016CQGra..33g5001L}
  Lynden-Bell, D. and Bi{\v c}{\'a}k, J.
\newblock {Pressure in Lema{\^i}tre-Tolman-Bondi solutions and cosmologies}.
\newblock {\em Classical and Quantum Gravity}. 33:075001, April 2016.  

\bibitem{2005JCAP...07..003M}
I.~{Maor} and O.~{Lahav}.
\newblock {On virialization with dark energy}.
\newblock {\em \jcap}, 7:003, July 2005.

\bibitem{2008PhRvD..78l3504U}
S.~{Unnikrishnan}, H.~K. {Jassal}, and T.~R. {Seshadri}.
\newblock {Scalar field dark energy perturbations and their scale dependence}.
\newblock {\em \prd}, 78(12):123504, December 2008.

\bibitem{2017JCAP...11..048B}
  R. C. Batista and V. Marra.
  \newblock {\em Clustering dark energy and halo abundances}.
  \newblock {\em \jcap}, 2017:48, November 2017.
  

\bibitem{2008PhRvD..77f7301A}
  L. R. Abramo, R. C. Batista, L. Liberato and R. Rosenfeld.
  \newblock {\em Dynamical mutation of dark energy}.
  \newblock {\em \prd}, 77:067301, March 2008.
  

\end{thebibliography}

\end{document}